\documentclass{pas}

\usepackage{aas-macros}
\usepackage{textcmds}
\usepackage{multirow}
\usepackage[authoryear]{natbib}
\usepackage{graphicx}
\usepackage{tikz}
\usepackage{nicefrac}
\usepackage{tabularx}
\usepackage{longtable}
\usepackage{nicefrac}
\usepackage{wasysym}
 \usepackage{adjustbox}
\begin{document}

\lefttitle{Publications of the Astronomical Society of Australia}
\righttitle{Sharma et al.}

\jnlPage{1}{15}
\jnlDoiYr{2026}
\doival{10.1017/pasa.xxxx.xx}

\articletitt{Research Paper}

\title{On the rates of binary star collisions around hypermassive black holes}

\author{
Megha Sharma,$^{1}$\thanks{E-mail: megha.sharma@monash.edu},
Alexander Heger$^{1}$, Evgeni Grishin$^{1,2}$, and Daniel J.\ Price$^{1}$}

\affil{$^1$School of Physics and Astronomy, Monash University, Wellington Rd, Clayton VIC 3800, Australia}
\corresp{M. Sharma, Email: megha.sharma@monash.edu}

\affil{$^2$OzGrav: Australian Research Council Centre of Excellence for Gravitational Wave Discovery}

\citeauth{Sharma M. et al. 2026 {\it Publications of the Astronomical Society of Australia} {\bf 00}, 1--15. https://doi.org/10.1017/pasa.xxxx.xx}

\history{(Received xx xx xxxx; revised xx xx xxxx; accepted xx xx xxxx)}

\begin{abstract}
Observations of the Milky Way show that galaxies have binary stars in their centres.  Using Monte Carlo simulations, we model the population statistics of encounters of such binary stars with galactic central black holes (BHs) of mass $10^5$--$10^{10}\,\mathrm{M}_\odot$ (hypermassive BHs $\geq 10^8\;\mathrm{M}_\odot$; HMBHs).  Whereas main-sequence single stars cannot encounter tidal disruption events around HMBHs, stellar collisions of main-sequence binary stars may produce observable transients.  We use the post-Newtonian Einstein--Infeld--Hoffmann (EIH) equations as implemented in the three-body code \textsc{Multistar}.  We simulate $100{,}000$ models each for six black hole masses and 10 different binary parameter distributions.  We find that stellar collisions occur in all models, with collision fractions ranging from $\sim 7\%$ for circular, equal-mass binaries to $\sim 0.2\%$ for the observationally motivated \citet[][MS17]{Moe2017} binary distribution.  Most collisions occur preferentially for prograde, tight binaries at low impact parameters $\beta \equiv r_\mathrm{t}/r_\mathrm{p}$, where the tidal force of the black hole drives up the binary eccentricity during pericentre passage.  Collision velocities are close to the binary escape velocity for lower-mass black holes ($\leq 10^8\;\mathrm{M}_\odot$), but span a wider range for higher-mass black holes, implying an increasing fraction of stellar mergers.  Using empirical loss-cone rates and the local black hole mass function, we estimate detectable LSST collision rates of $\sim8\mathord,600$ and $\sim3\mathord,000$ events per year for our Kroupa mass function, \"{O}pik's law, inclined orbit model and our MS17 model, respectively, around $10^6\;\mathrm{M}_\odot$ black holes; $\sim 330$ and $\sim 150$ events per year around $10^8\;\mathrm{M}_\odot$ black holes; and $\sim 3.5$ and $\sim 2.8$ events per year around $10^9\;\mathrm{M}_\odot$ black holes.  These results suggest that binary stellar collisions around HMBHs represent a potentially observable population of nuclear transients that may contribute to the growing sample of ambiguous nuclear transients detected by wide-field optical surveys.
\end{abstract}

\begin{keywords}
(stars:) binaries: general, Galaxy: nucleus, black hole physics , chaos
\end{keywords}

\maketitle

\section{Introduction}

Most galaxies have a supermassive black hole (SMBH) at their centre \citep{Kormendy2013}.  These black holes (SMBH masses $\leq 10^8\;\mathrm{M}_\odot$) commonly coexist with nuclear star clusters --- dense, compact stellar systems that dominate the light in the central few parsecs \citep{Graham2009,Georgiev2016,Neumayer2020}. As the black hole mass increases, however, its stronger gravitational potential can inhibit the formation and survival of a nuclear star cluster \citep{Georgiev2016}. Despite this, disc-like stellar structures may still form around such black holes \citep{Arca2017}. We define black holes with $M_\bullet \geq 10^8\;\mathrm{M}_\odot$ as \textit{hypermassive black holes} (HMBHs).

A star can be tidally disrupted when the gravitational forces of the black hole overcome the self-gravity of the star \citep{Hills1975} --- a tidal disruption event (TDE). TDEs may dominate the growth of intermediate-mass black holes \citep{Rizzuto2023}, potentially  explaining observations of black holes with $M_\bullet \geq 10^9\,\mathrm{M}_\odot$ \citep{Wang2021,Farina2022,Greene2024}.

%\eg{[The next two sentences are unclear. Let me know if you'd like me to rephrase it.]} Main-sequence stars would be directly swallowed by 
For black holes with $M_\bullet \gtrsim 10^{7.5}\,\mathrm{M}_\odot$, main sequence stars can no longer be tidally disrupted because the tidal radius becomes smaller than the Schwarzschild radius \citep{vanVelzen2018,Polkas2023}. However, a spinning black hole could still induce a TDE \citep{Kesden2012tidal,Mummery2024}. TDE candidates around HMBHs remain rare, with only a few reported cases \citep{Leloudas2016,Gezari2021,Graham2026} found for black holes $> 10^8\;\mathrm{M}_\odot$.

Yet, not all transient flares need to originate from TDEs. Stellar collisions could produce similar observational signatures \citep{Ann2021}. Despite the seeming lack of a nuclear star cluster around high mass black holes \citep{Antonini2013,Georgiev2016,Hoyer2024}, stars could still collide in such environments, e.g. from dense stellar discs \citep{Tremaine1995,Arca2017}.  In this work, we aim to quantify the collision rate for binary stars around HMBHs. 

High-mass stars are more likely to be binary or in multiple systems than low-mass stars. The multiplicity fraction for $M_* \geq 16\;\mathrm{M}_\odot$ is $\geq 80\%$ \citep{Duchene2013}. In contrast, for $M_* \leq 0.1\;\mathrm{M}_\odot$ the fraction is only $22\%$ for main-sequence stars. About $75\%$ of O-type stars have a companion. About $70\%$ of B- and A-type stars are part of a multiple-star system \citep{Raghavan2010,Offner2023}.

Our Galactic Centre also hosts binaries, including IRS~16SW \citep{Ott1999}, IRS~16NE, E60/S4-258 \citep{Pfuhl2014}, S2-36 \citep{Gautam2019,Gautam2024}, and D9 \citep{Peiker2024}. The estimated binary fraction of young, massive stars in the Galactic Centre is $\sim 72\%$ or $\sim 42\%$ within one and two standard deviations, respectively. This is consistent with local binary fractions \citep{Gautam2024}, but higher than the predictions of \citet{Chu2023}. 

A star close to its tidal radius around a $10^9M_\bullet$ HMBH can have at Keplerian velocities of $\sim 30\%$ of the speed of light. \citet{Metzger2017} proposed that when an extreme mass ratio inspiral (EMRI) --- a star and a black hole system collides with another star, it might result in a TDE impostor. \citet{Mastrobuono2021} performed $N$-body simulations to understand the collision rates of stars in the Galactic Centre. \citet{Rose2024} and \citet{Rose2025} used a semi-analytical model to study the effect of collisions on the stellar density profile in the Galactic Centre, and proposed that such events could result in stars entering the tidal radius and being disrupted by the black hole. Stellar collisions have been proposed to explain the G-objects --- dusty yet dynamically stellar objects --- in the Galactic Centre \citep{Witzel2014,Stephan2016,Rose2023,Gibson2025}.

To understand the dynamics of stars around SMBH, \citet{Dodici2025} solved the secular equations of motion of a hierarchical three-body system. They found that binaries would be on tight orbits around the SMBHs due to diffusive excitation of stellar tides, which is relevant when pericenter separations are smaller than a few stellar radii. \citet{Marklund2025} used a $N$-body code to solve this three-body problem, where the SMBH's effect is implemented in a secular manner. They argued that most binary star systems might have originated closer to the SMBH compared to their present location. \citet{Mcguire2026} used a semi-analytical model and found that general relativistic precession reduces the von-Zeipel-Lidov-Kozai (ZLK) effect \citep{Naoz2016}, resulting in fewer mergers than with Newtonian physics.

 The collision of high-velocity stars has been proposed as an event that can mimic TDEs or result in a transient \citep{Balberg2013,Yu2024}. These works assume collisions between two unbound stars in the Nuclear Star Cluster that collide at high velocities, exceeding their stellar binding energy, potentially resembling supernovae \citep{Balberg2023,Balberg2024,Seoane2023}. The ejecta of these events could fall back onto the black hole, and the accretion of this material onto the black hole could result in a transient that might resemble a TDE \citep{Ryu2024,Ryu2025}. All the studies listed above focus on collisions around a black hole with a mass similar to that of the Galactic Centre black hole.

\citet{Yu2024} and \citet{Hu2024} considered binary stars encounters around $10^6\;\mathrm{M}_\odot$ black hole. They used the Newtonian code \textsc{Rebound}, whereas we use a post-Newtonian (PN) code for this work. They argued that these encounters could result in exotic events, which they followed up on using Smoothed Particle Hydrodynamics (SPH) simulations in \citet{Yu2025}.

In this paper, we explore the collisions of binary stars around black holes. Specifically, we estimate the rate of collisions of binary star components with each other, caused by the gravitational forces of the black hole on the binary orbit using the Monte Carlo code \textsc{Multistar}\footnote{\url{https://2sn.org/multistar/doc/index.html}}, for our three-body simulations. We use the Einstein-Infeld-Hoffman equations \citep*{Einstein1938} instead of pair-wise PN terms, because in our previous work we found that the pair-wise method can result in non-physical decrease in separation due to not considering the cross terms \citep{Sharma2026}. We consider a range of black hole masses from $10^5\;\mathrm{M}_\odot$ to $10^{10}\;\mathrm{M}_\odot$. Our main focus is on HMBHs. Binary star systems are likely to exist around HMBHs as well, and hence, understanding whether collision mechanism can result in any observational signatures that may resemble TDEs is important.

% \eg{[you should find an appropriate place to cite our previous benchmark paper, either in the intro of the very beginning of sec 2. This is the main motivation why we choose to use he EIH method so it's puzzling why it's not mentioned. Also I think the benchmark paper is the first we use multistar, so maybe cite it whenever multistar is mentioned at first instance.]}

Our paper is structured as follows: Section~\ref{sec:methods_6} describes our methods, Section~\ref{sec:results_6} presents the results, we discuss them in Section~\ref{sec:discussion_6}, and we conclude in Section~\ref{sec:conclusion_6}.

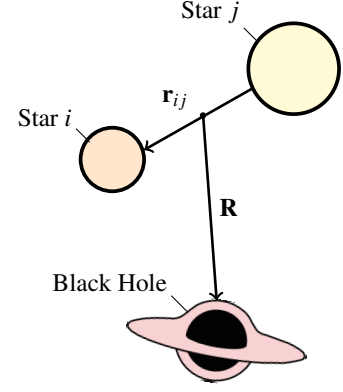
\begin{figure}
\centering
\begin{tikzpicture}[scale=0.6, transform shape]

% planet image
% star circle
\draw[line width=1.4pt, transform shape=false,draw=black, fill=orange!20] (-2.3,4) circle [radius=0.70];
% \node at (-2.3,4) 

% {\includegraphics[width=1.2cm]{planete_seule.png}};

% star circle
\draw[line width=1.4pt, transform shape=false,draw=black, fill=yellow!20] (1.7,6) circle [radius=1];

% BH image
\node[anchor=center, rotate=-10] at (0,0) {\includegraphics[width=4cm]{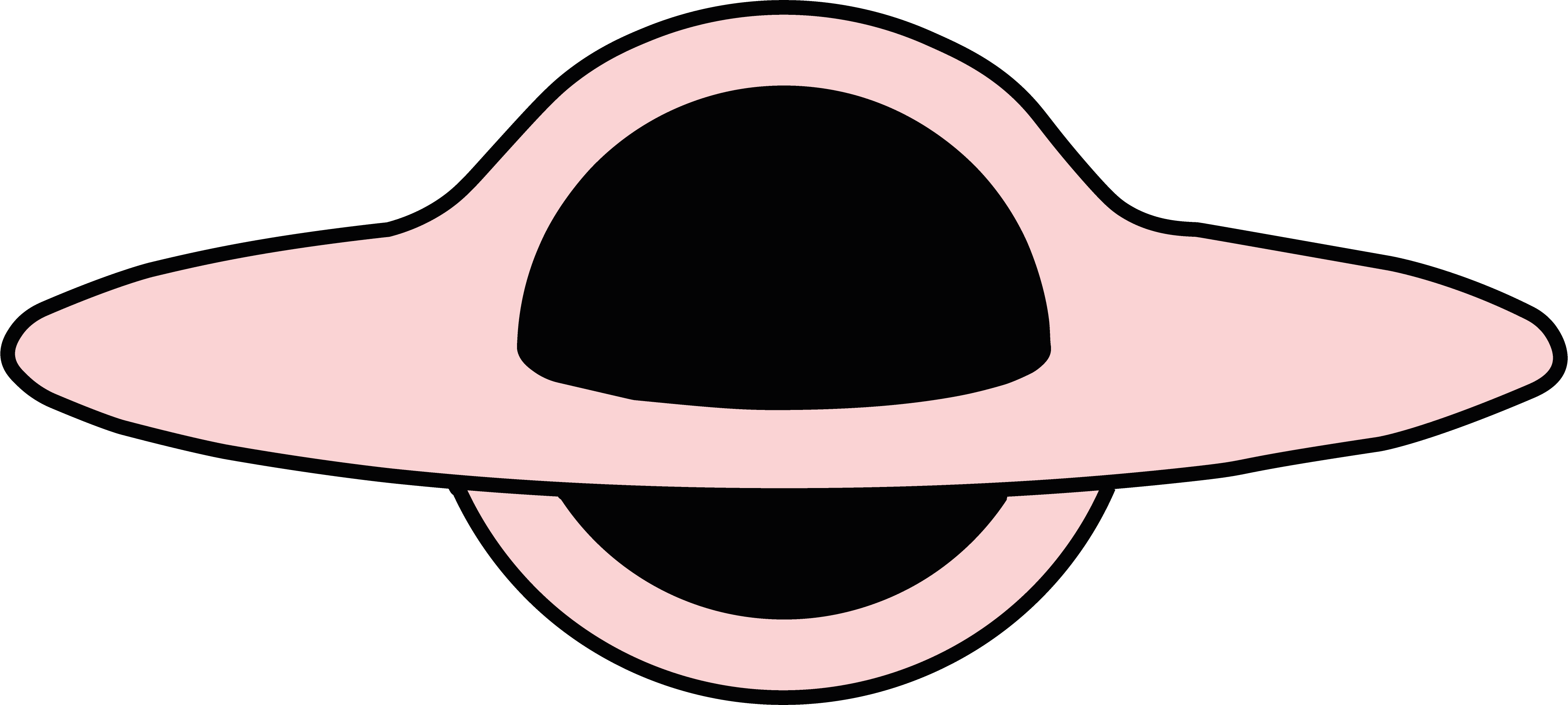}};

% r
\draw[->, line width=1pt, transform shape=false] (0.78,5.56) -- (-1.6,4.21)
node[midway, above, anchor=south east] {$\mathbf{r}_{ij}$};

% R
\draw[->, line width=1pt, transform shape=false] (-0.3,5.0) -- (0,0.9)
node[midway, right] {$\mathbf{R}$};

% cm
\fill[transform shape=false] (-0.3,4.97) circle [radius=2pt,draw=black, fill=red];

% text labels (fixed size)
\node[scale=1, transform shape=false, anchor=south east, align=right] at (-3,4.53) {Star $i$};
\draw[transform shape=false] (-2.8,4.38) -- (-3.1,4.68);

\node[scale=1, transform shape=false, anchor=south east, align=right] at (0.7,6.8) {Star $j$};
\draw[transform shape=false] (0.9,6.6) -- (0.6,6.9);

\node[scale=1, transform shape=false, anchor=south east, align=right] at (-0.9,0.9) {Black Hole};
\draw[transform shape=false] (-0.65,0.65) -- (-1.,1.);

\end{tikzpicture}

\caption{Jacobi coordinates, where $\mathbf{r}_{ij}$ is defined from star $j$ to star $i$, and $\mathbf{R}$ from the system's centre of mass to the black hole (we thank Gauthier Leurent for helping with this diagram).}
\label{fig:jacobi_coords}
\end{figure}

\section{Methods}
\label{sec:methods_6}

\begin{table*}
\centering
\caption{Possible outcomes of a binary–black hole encounter.}
\label{tab:outcomes}
\begin{tabularx}{\textwidth}{lX}
\hline
\textbf{Outcome} & \textbf{Description} \\
\hline
Collision & Two stars collide with each other. Using the radius of the stars, if the separation between them is less than the sum of their radii, then we consider the stars to have collided. \\
Tidal Disruption Event (TDE) & One or both stars are tidally disrupted by the black hole. \\
Swallow & One or both stars are directly swallowed by the black hole without a TDE. \\
Capture & At least one star is captured into a bound orbit around the black hole. \\
Escape & Binary star escapes undisrupted on a hyperbolic orbit. \\
Disruption & Both stars escape on a hyperbolic orbit but are no longer bound to each other. \\
\hline
\end{tabularx}
\end{table*}

We use the in-house dynamics code \textsc{Multistar}, which solves the Einstein--Infeld--Hoffmann equations \citep{Einstein1938}, 
\begin{align}
\mathbf{a}_i &= 
 -\sum_{j\ne i} \frac{G m_j\, \mathbf{r}_{ij}}{r_{ij}^3}
+ \frac{1}{c^2} \sum_{j\ne i} 
\frac{G m_j\, \mathbf{r}_{ij}}{r_{ij}^3}
\Bigg[
4\,\frac{G m_j}{r_{ij}}
+ 5\,\frac{G m_i}{r_{ij}} \nonumber\\
&\quad+ \sum_{k\ne i,j} \frac{G m_k}{r_{jk}}
+ 4 \sum_{k\ne i,j} \frac{G m_k}{r_{ik}}
- \frac{1}{2} \sum_{k\ne i,j} 
\frac{G m_k}{r_{jk}^3} (\mathbf{r}_{ij}\!\cdot\!\mathbf{r}_{jk}) \nonumber\\
&\quad- v_i^2 + 4\,\mathbf{v}_i\!\cdot\!\mathbf{v}_j
- 2 v_j^2
+ \frac{3}{2}(\mathbf{v}_j\!\cdot\!\mathbf{n}_{ij})^2
\Bigg] \nonumber\\
&\quad - \frac{7}{2c^2}
\sum_{j\ne i} \frac{G m_j}{r_{ij}}
\sum_{k\ne i,j} \frac{G m_k\, \mathbf{r}_{jk}}{r_{jk}^3} \nonumber\\
&\quad + \frac{1}{c^2}\sum_{j\ne i} \frac{G m_j}{r_{ij}^3}
\mathbf{r}_{ij}\cdot(4\mathbf{v}_i - 3\mathbf{v}_j)\mathbf{v}_{ij}\;,
\label{eq:EIH}
\end{align}
where $m$ is the mass, $\mathbf{r}$ is the position vector, $\mathbf{v}$ is the velocity vector, $\mathbf{r}_{ij} = \mathbf{r}_i-\mathbf{r}_j$, and $\mathbf{n}_{ij} = \mathbf{r}_{ij}/r_{ij}$. It
corresponds to the first-order post-Newtonian corrections. These equations have been used previously to solve the orbits of the S-star cluster around Sagittarius~A* \citep{Parsa2017}, though only for two-body rather than three-body systems. \citet{Suzuki2021} solved the EIH equations for a three-body system, considering interactions involving black holes of up to intermediate mass. We have extended this method to hypermassive black holes.

\textsc{Multistar} uses Jacobi coordinates as shown in Figure~\ref{fig:jacobi_coords} and uses a Bulirsch–Stoer integrator \citep{Bulirsch1966}. We treat our stars and black hole as point particles, i.e., we ignore tidal interactions and spin.  We note that we calculate the radius of our stars using \citep{Salaris2005}
\begin{align}
    R_* = \mathrm{R}_\odot \left\{\begin{array}{lll}\left( \frac{M_*}{\mathrm{M}_\odot}\right)^{0.8}&\phantom{xxx}& \mathrm{for}\; M_* < \mathrm{M}_\odot\;,\\
        \left( \frac{M_*}{\mathrm{M}_\odot}\right)^{0.65} &\phantom{xxx}& \mathrm{otherwise}\;,\end{array}\right.  
        \label{eq:stellar_radius}
\end{align}
where $M_*$ and $R_*$ are mass and radius of the star.
We place the binary systems on zero-energy trajectories (parabolic) around the black hole. We first determine the velocity of the centre of mass of the binary and then place the binary at $100\;r_\mathrm{t}$ from the black hole. We note that $r_\mathrm{t}$ (tidal radius), given by
\begin{equation}
    r_\mathrm{t} = a_\mathrm{b} \left(\frac{M_\bullet}{m_\mathrm{b}} \right)^{1/3}\;,
\end{equation}
where $a_\mathrm{b}$ is the semi-major axis of the binary, $M_\bullet$ is the mass of the black hole, and $m_\mathrm{b}$ is the mass of the binary, can be close to the innermost stable circular orbit (ISCO) as the black hole mass increases. We therefore limit our pericentre to outside this distance. We use bisection to determine the orbit with the desired pericentre, $r_\mathrm{p}$, ensuring that the same impact parameter, $\beta \equiv r_\mathrm{t}/r_\mathrm{p}$ is compared across codes, even in the general relativistic regime.

We classify the interaction of a binary with a black hole into different outcomes as listed in Table~\ref{tab:outcomes}. These include collision of stars which is the main focus of this paper. If the separation between the two stars is less than the sum of their radii, we classify them as a collision. One or both stars can end up captured by the black hole, where the other star escapes. This is known as the Hills mechanism \citep{Hills1988}. Other outcomes include disruptions which affect the stellar orbits, or a TDE. 

\subsection{Model parameters}
We vary a range of parameters for our simulations. Table~\ref{tab:all_models} lists the various scenarios tested in this paper. We consider a fiducial case where the two stars of the binary are on a circular orbit about each other, with a semi-major axis of $1\;\mathrm{au}$. We only vary the mean anomaly, and consider a penetration factor $\beta \equiv r_\mathrm{t}/r_\mathrm{p} = 1$. For simplicity, we consider stars with a mass of $1\;\mathrm{M}_\odot$. We then vary different properties of the binary system --- such as the semi-major axis, eccentricity, inclination, and mass of the binary stars --- to explore how these parameters affect the number of collisions.

Observations of binary stars show that eccentricity and period are related such that binaries with periods $< 10$ days are circular, while those above are dynamically relaxed such that their distribution is close to the thermal distribution $p(e) = 2e$ \citep{Ambartsumian1937,Duquennoy1991,Raghavan2010}. \citet{Duchene2013} argue that the distribution is closer to a Gaussian. Although the validity of both distributions is limited by the small sample size, for this work we adopt a Gaussian with a mean of $0.3$ and a standard deviation of $0.15$ for binaries with periods $\geq 10$ days, with a cut-off at $0$ and $0.9$ \citep{Stepinkski2001,Mandel2015}. To ensure that the binary pericentre around the black hole never falls inside the ISCO, we set a lower limit on the semi-major axis of
\begin{equation}
    a_\mathrm{min} = 6\, r_\mathrm{g}\, \beta \left(\frac{m_\mathrm{b}}{M_\bullet}\right)^{\!\nicefrac{1}{3}}\;,
\end{equation}
where $r_\mathrm{g} \equiv G M_\bullet/c^2$. We set the minimum orbital period to $P_{\rm min} = \max\left[P(a_{\rm min}),\, 1000~{\rm d}\right]$, where $P(a_{\rm min})$ is the period corresponding to $a_{\rm min}$. If the period corresponding to $a_\mathrm{min}$ exceeds $1{,}000$ days, we set the maximum period to $P_\mathrm{min} + 1{,}000$ days.

To understand the effect of the semi-major axis on the outcomes, we consider a distribution $p(a) \propto a^{-1}$, also known as \"{O}pik's law \citep{Opik1924}. We also vary the inclination by drawing it from a cosine distribution, such that $i > 0$ corresponds to prograde and $i < 0$ to retrograde motion of the binary. We also adopt a distribution $p(\beta) \propto \beta^{-2}$ for the penetration factor (see Appendix~\ref{app:beta_dis}), where $\beta$ ranges from $0.01$ to $10$.

We consider different binary masses. We first consider a $1+1\;\mathrm{M}_\odot$ binary, followed by $10+10\;\mathrm{M}_\odot$ and $10+3\;\mathrm{M}_\odot$ binaries. We also consider the Kroupa mass function (KMF; \citealt{Kroupa2001}) to generate the mass of the primary star ($m_1$). The secondary mass ($m_2$) is drawn from a distribution $p(q) \propto q^{-3/4}$, where $q = m_2/m_1$ and $q \in [0.2,1]$, similar to \citet{Mandel2015}. We use a minimum and maximum mass of $0.1\;\mathrm{M}_\odot$ and $150\;\mathrm{M}_\odot$, respectively. Unlike \citet{Mandel2015}, however, we draw our semi-major axis from \"{O}pik's law, whereas they derived the distribution of binary semi-major axes that would undergo double TDEs --- which is not the primary focus of this work. Finally, we consider the \citet[][MS17]{Moe2017} binary star distribution, where the mass of the primary star lies within $0.3$--$50\;\mathrm{M}_\odot$. MS17 calculated the distribution of eccentricity, period, and mass ratio of binary stars of O and B-type and solar-type stars. To generate systems satisfying the condition that all models are outside the ISCO, we sample until the condition is met, which implies that the resulting distribution may be slightly biased. We generate $100{,}000$ simulations for each black hole mass, giving a total of $6$ million simulations. 
\begin{figure*}
    \centering
\includegraphics[width=\textwidth,height=\textheight,keepaspectratio]{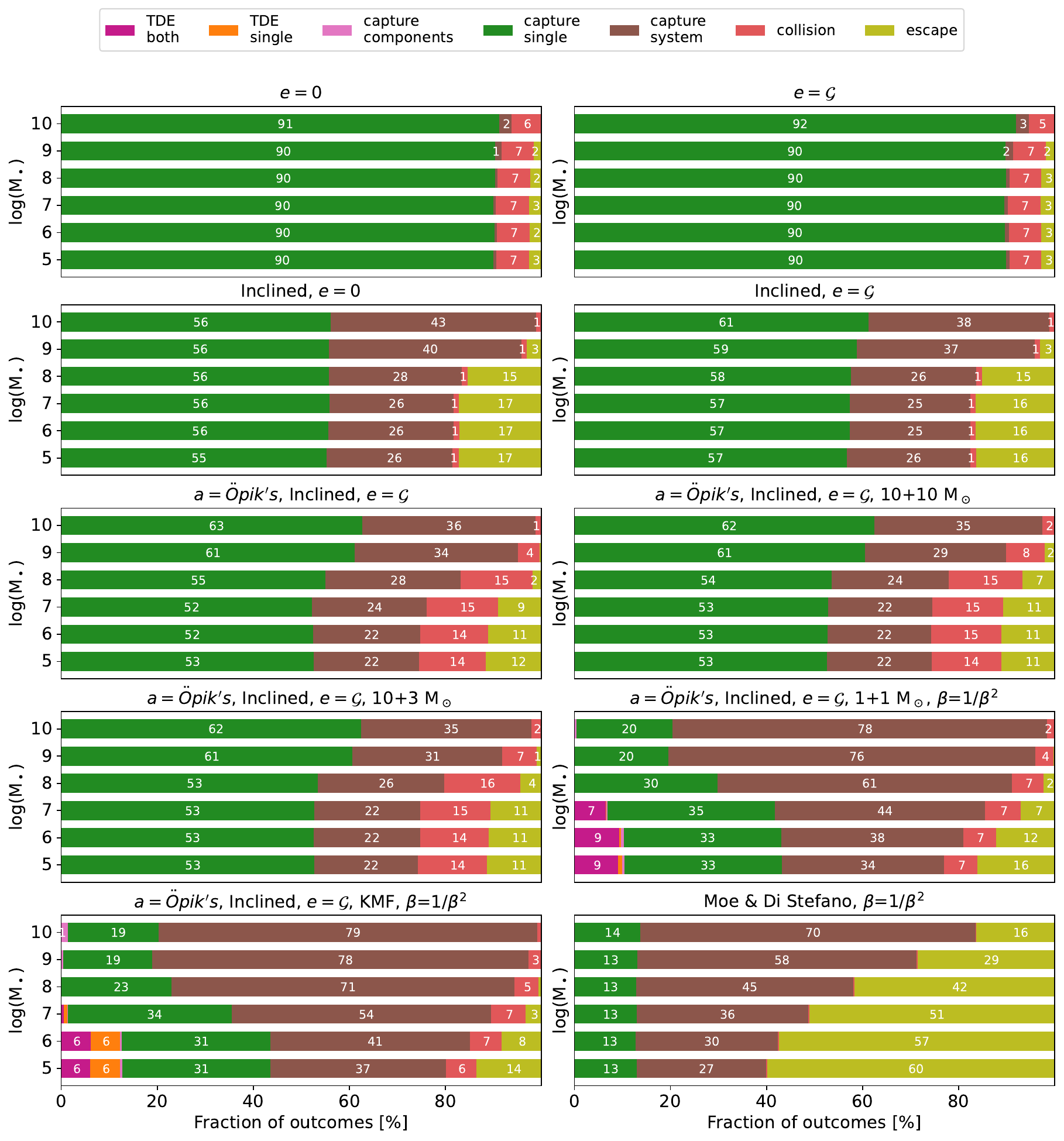}
    \caption{Different outcomes for different setups listed in Table~\ref{tab:all_models}. The most common outcome is primary or secondary capture while the other star escapes \citep{Hills1988}, capture of binary and escape. Double TDEs and primary or secondary star TDE takes place in our simulations where we use a $\beta$ distribution of $1/\beta^2$. Collision of stars can take place around HMBHs, but for MS17 only $\sim 0.2\%$ of binaries collide across different black holes.}
    \label{fig:collisions_cases}
\end{figure*}
\begin{figure*}
    \centering
\includegraphics[width=\textwidth,height=\textheight,keepaspectratio]{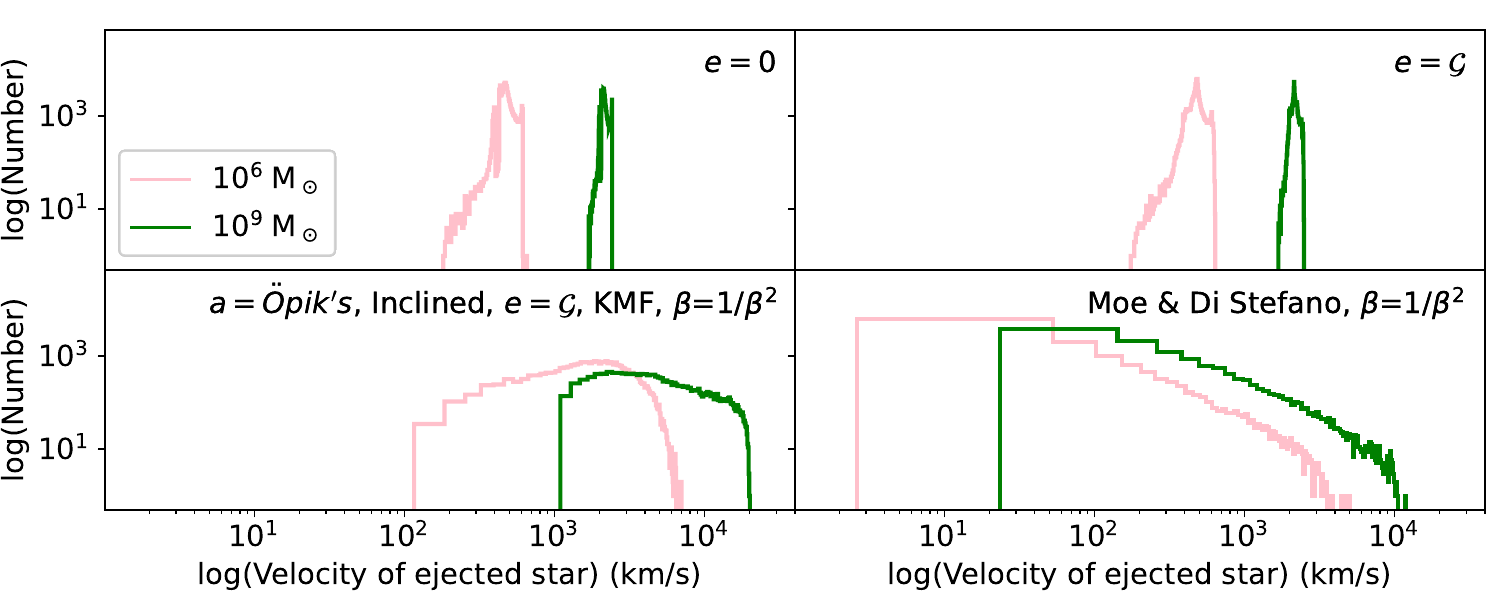}
    \caption{Ejected star velocity for $10^6\;\mathrm{M}_\odot$ and $10^9\;\mathrm{M}_\odot$ black holes. Most stars are ejected with velocities ($\sim 10^{2-4}\;\mathrm{km\;s^{-1}}$). }
    \label{fig:eject_vel}
\end{figure*}

\begin{figure*}
    \centering
\includegraphics[width=\textwidth,height=\textheight,keepaspectratio]{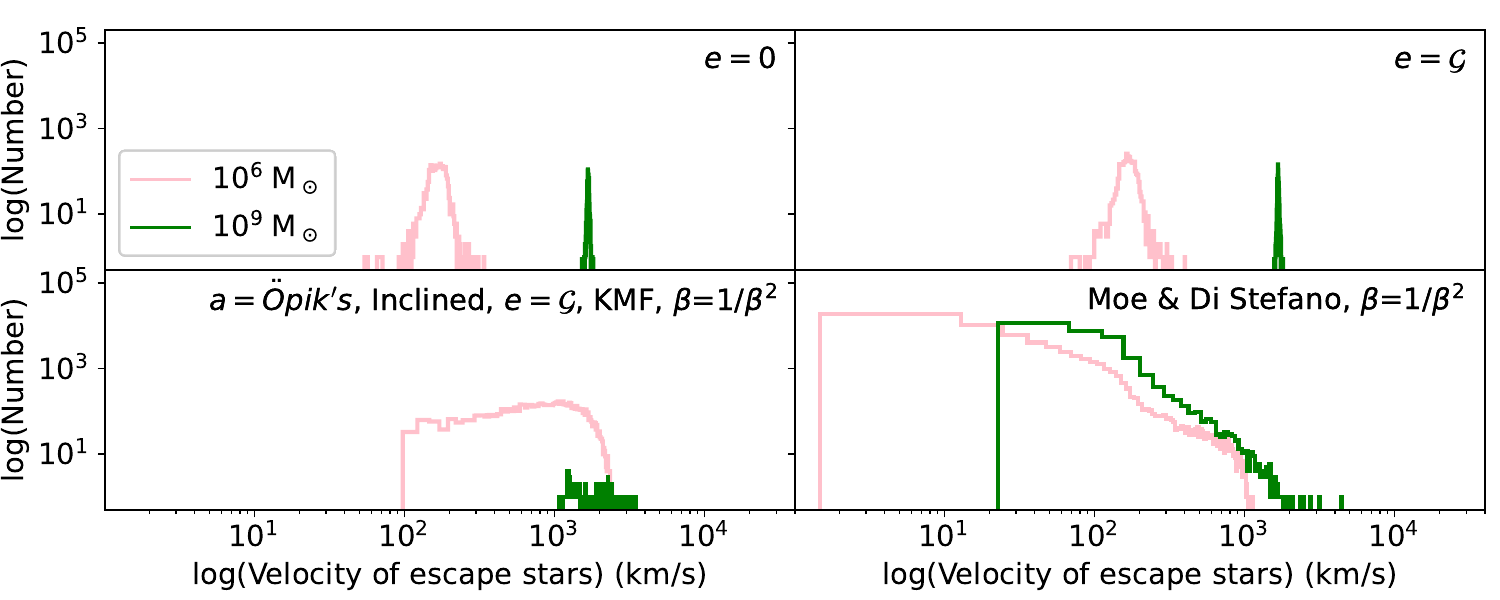}
    \caption{Escape velocity of binary stars that end up on hyperbolic orbits for $10^6\;\mathrm{M}_\odot$ and $10^9\;\mathrm{M}_\odot$ black holes.  }
    \label{fig:eject_vel_escape}
\end{figure*}

\begin{figure}
    \centering
    \includegraphics[width=\linewidth]{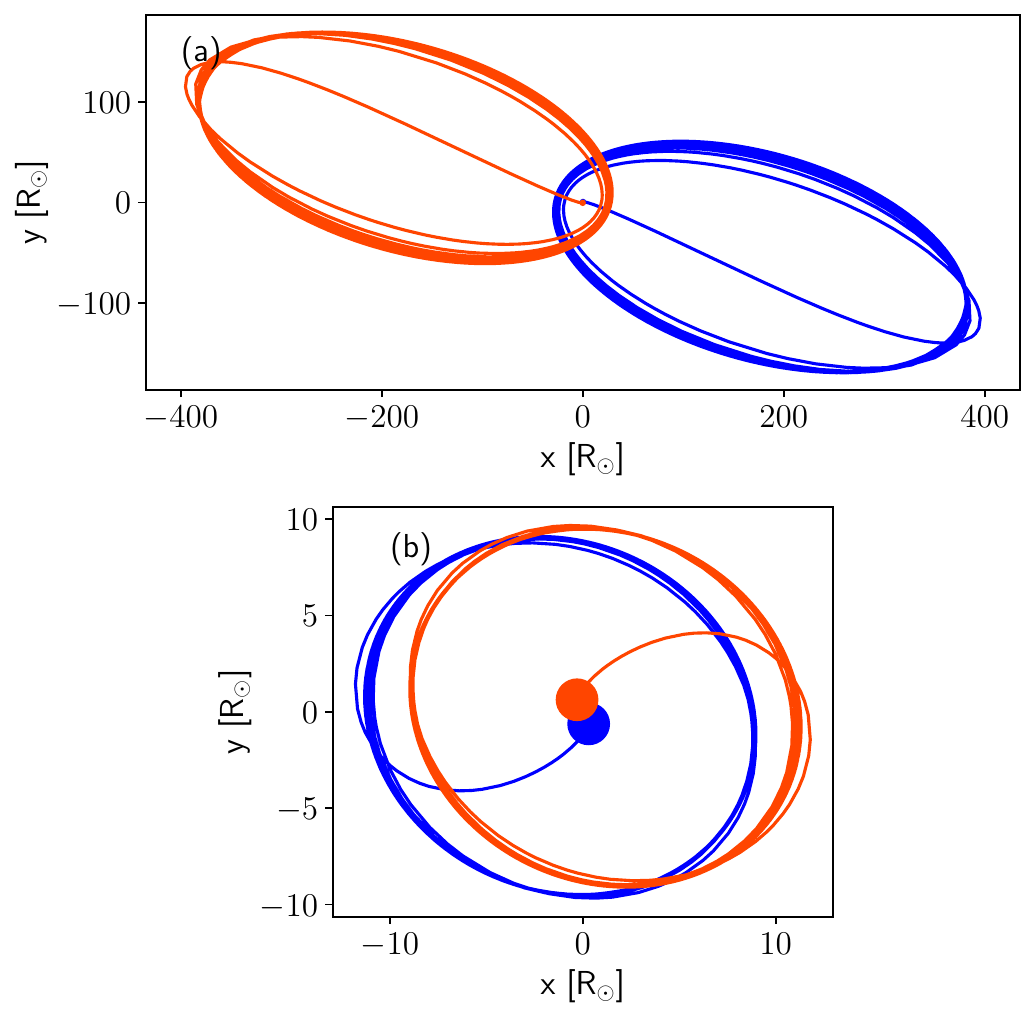}
    \caption{ Orbital evolution in the centre of mass frame of the binary stars. Red and blue line represent two different stars. Both the eccentric and circular orbits eventually collide due to the gravitational forces of the black hole. The initial parameters of the binary are generated from the MS17 distribution. }
    \label{fig:collision_example}
\end{figure}
\begin{figure*}
    \centering
    \includegraphics[width=\textwidth,height=\textheight,keepaspectratio]{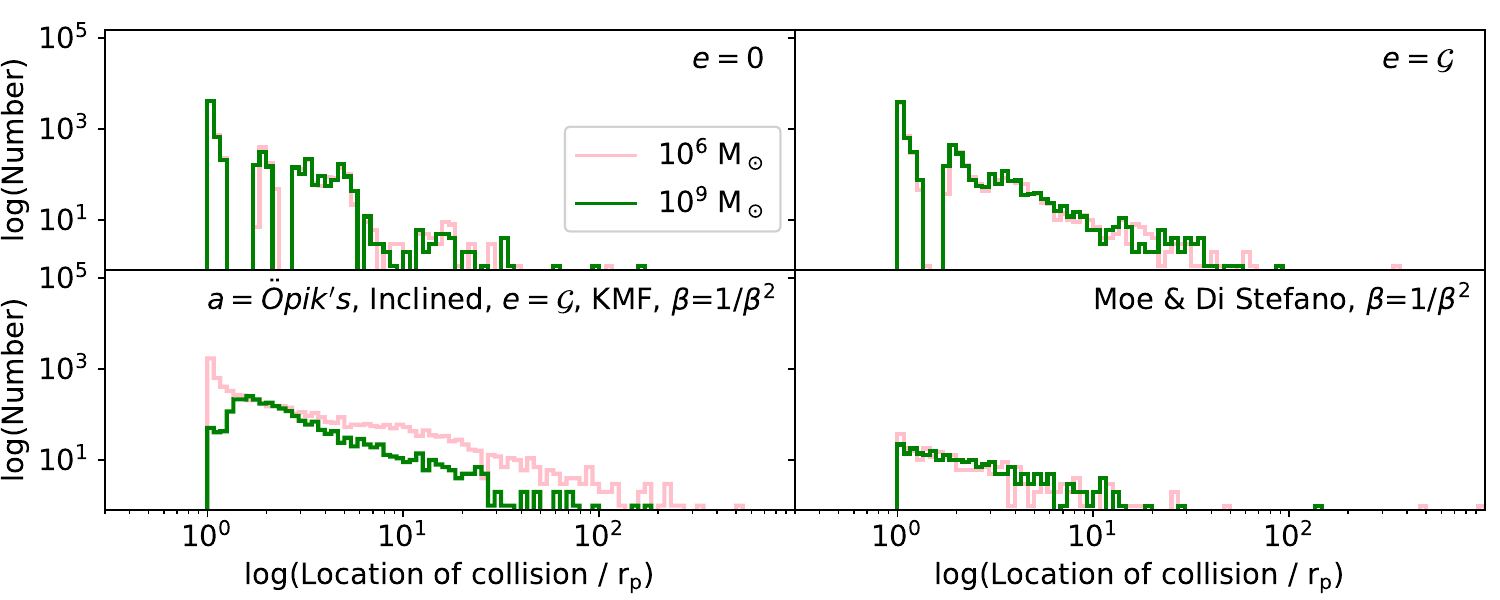}
    \caption{Binned collision locations relative to the pericentre of the orbit around the black hole, for $10^6\;\mathrm{M}_\odot$ and $10^9\;\mathrm{M}_\odot$ black holes. Each panel corresponds to models 1, 2, 9 and 10 from Table~\ref{tab:all_models}, showing the full parameter study. Most collisions occur close to pericentre; gaps in the distribution correspond to binaries that narrowly avoid collision at first pericentre passage and collide on a subsequent orbit.  }
    \label{fig:collision_loc}
\end{figure*}
\begin{figure*}
    \centering
\includegraphics[width=\textwidth,height=\textheight,keepaspectratio]{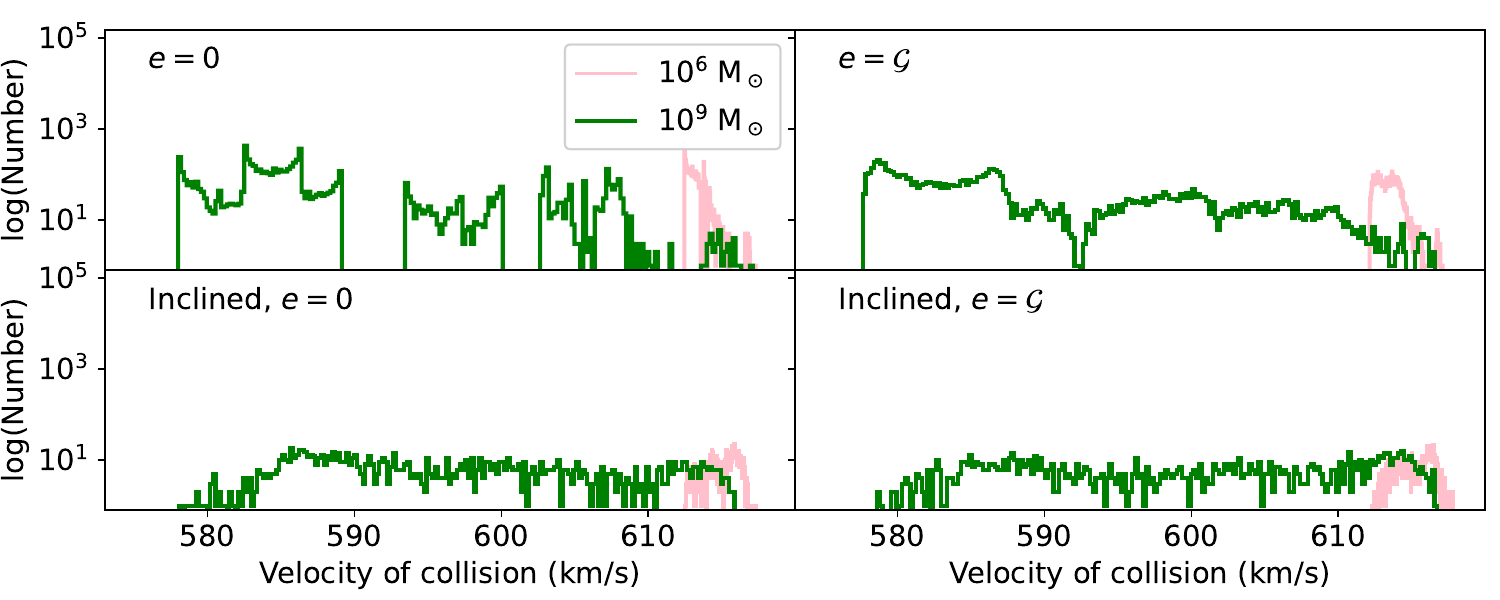}
    \caption{Relative velocity of colliding stars at the point of collision, for $10^6\;\mathrm{M}_\odot$ 
and $10^9\;\mathrm{M}_\odot$ black holes, showing the four models with a fixed 
semi-major axis of $1\;\mathrm{au}$ (Models~1--4 in Table~\ref{tab:all_models}). 
For lower-mass black holes, collision velocities are tightly clustered around the 
binary escape velocity (${\sim}617\;\mathrm{km\,s}^{-1}$ for a $1+1\;\mathrm{M}_\odot$ 
binary at contact), indicating that the binary's internal dynamics dominate. For 
higher-mass black holes, the distribution broadens as the black hole's stronger 
gravitational influence drives stars together across a wider range of orbital 
configurations, implying an increasing fraction of mergers rather than grazing 
collisions.  }
    \label{fig:collision_vel}
\end{figure*}
\begin{figure*}
    \centering
\includegraphics[width=\textwidth,height=\textheight,keepaspectratio]{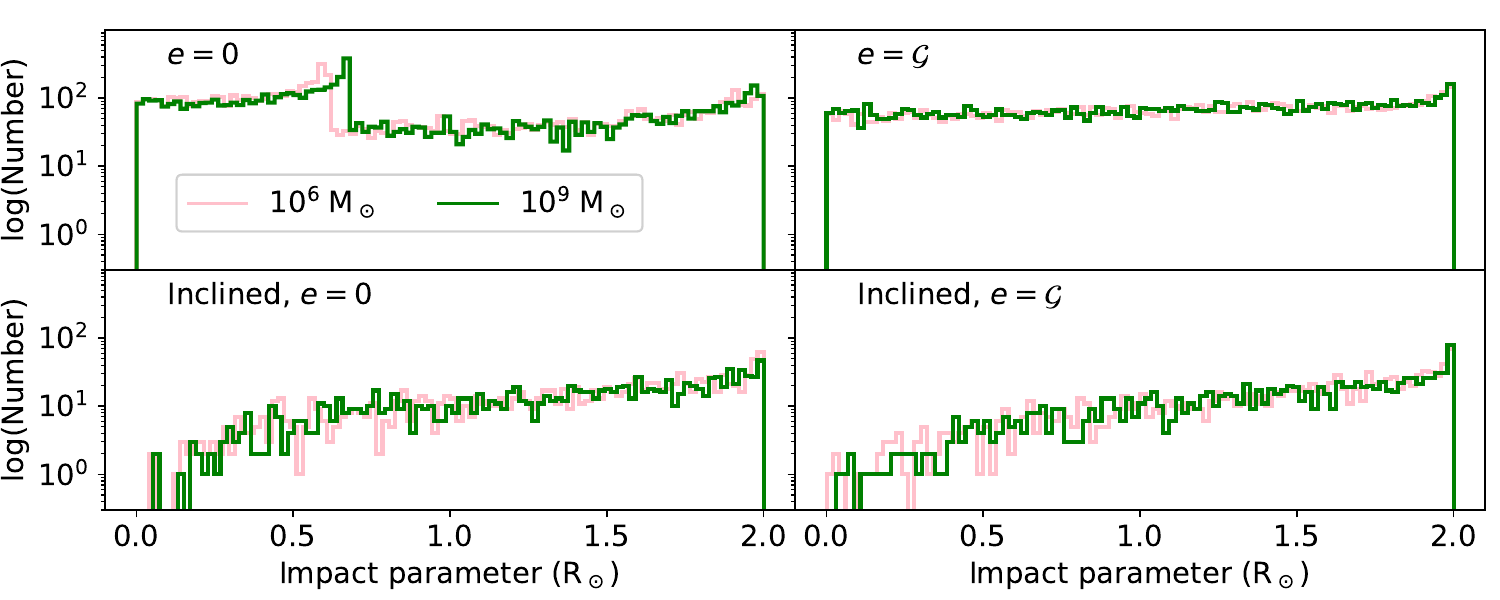}
    \caption{Impact parameter at the point of collision for the four models considering only 
circular $1+1\;\mathrm{M}_\odot$ binaries with a semi-major axis of $1\;\mathrm{au}$ 
(Models 1--4 in Table~\ref{tab:all_models}). For non-inclined models (Models~1 
and~2), the impact parameter is approximately uniformly distributed, implying 
collisions ranging from head-on to grazing. For inclined models (Models~3 and~4), 
grazing encounters are more common, with the collision rate increasing with impact 
parameter.}
    \label{fig:collision_b}
\end{figure*}
\begin{figure*}
    \centering
\includegraphics[width=\textwidth,height=\textheight,keepaspectratio]{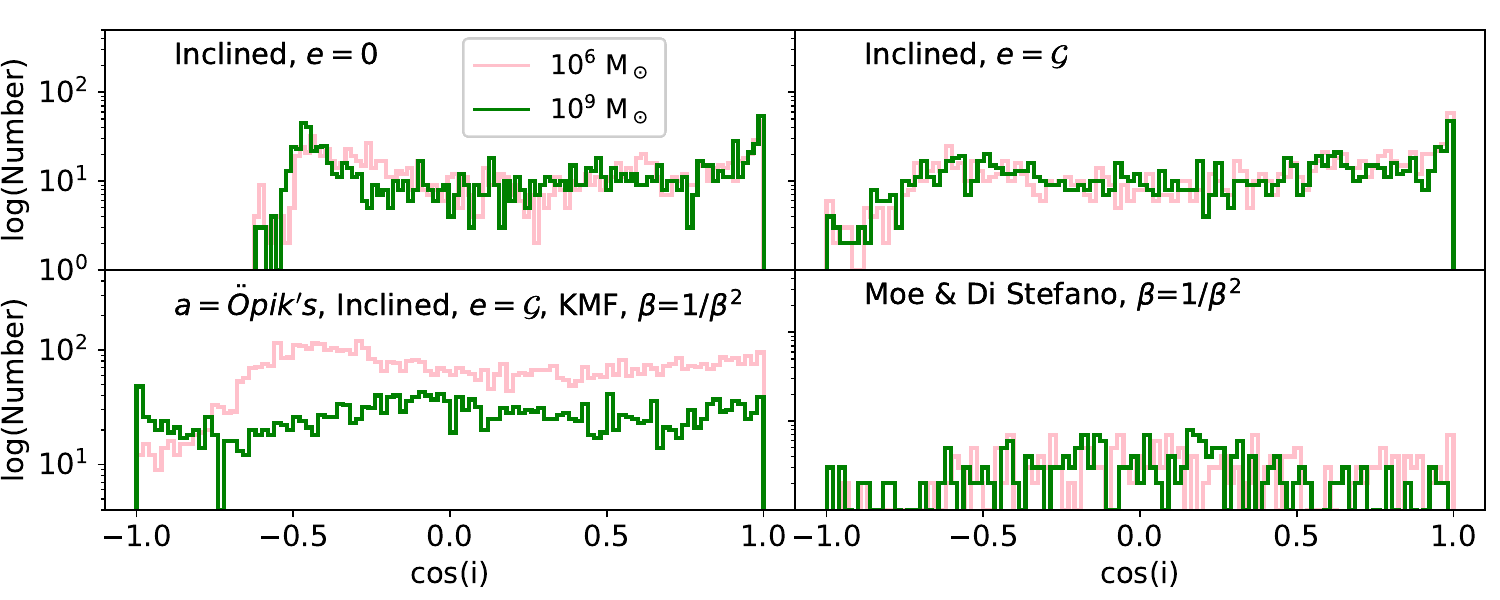}
    \caption{ Distribution of binary inclinations for colliding systems, for $10^6\;\mathrm{M}_\odot$ 
and $10^9\;\mathrm{M}_\odot$ black holes. Prograde binaries ($\cos i > 0$) collide 
more frequently than retrograde ones ($\cos i < 0$), with no collisions occurring 
for nearly retrograde, circular inclined orbits ($\cos i \lesssim -0.5$). This preference arises from near-resonance between the binary's orbital 
motion and the gravitational forces of the black hole, analogous to the prograde enhancement 
seen in galactic disc interactions \citep{Toomre1972, Donghia2010}.}
    \label{fig:collision_inc}
\end{figure*}

\begin{figure}
    \centering
\includegraphics[width=\linewidth]{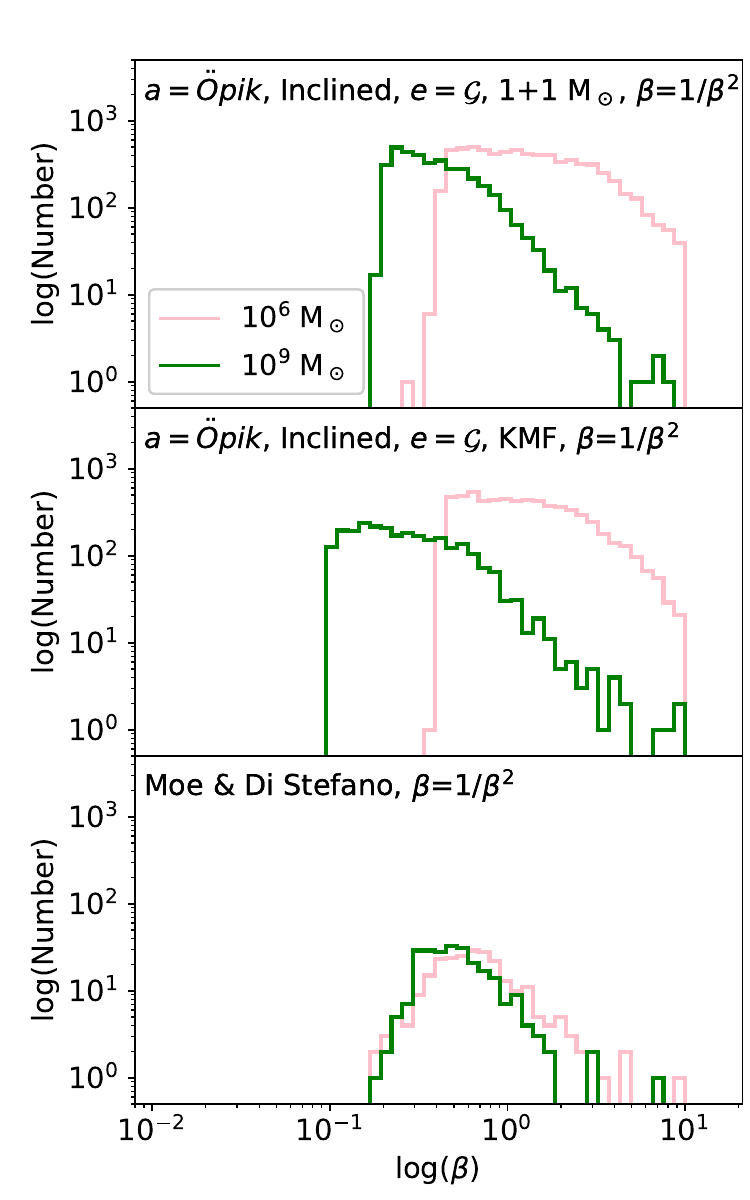}
    \caption{Distribution of impact parameters $\beta \equiv r_\mathrm{t}/r_\mathrm{p}$ for 
colliding binaries in Models~8, 9, and~10, for $10^6\;\mathrm{M}_\odot$ and 
$10^9\;\mathrm{M}_\odot$ black holes. Most collisions occur at low $\beta$, 
particularly for higher-mass black holes, where the gravitational interaction with 
the black hole during pericentre passage drives up the binary eccentricity, causing 
the stars to collide. This effect is more pronounced for tight binaries, which 
complete more orbits during a single pericentre passage, allowing coherent growth 
of the binary eccentricity until collision occurs. }
    \label{fig:collision_beta}
\end{figure}

\begin{table*}
    \centering
    \caption{Parameters used for each model. Column $1$ lists the model number,
    followed by eccentricity ($e$) and semi-major axis ($a$) of the binary stars orbit.
    Column $4$--$7$ list the inclination ($i$), mean anomaly ($M$), longitude of
    ascending node ($\Omega$), and argument of periapsis ($\omega$). Followed by columns listing
    the mass of binary ($m_\mathrm{b}$), impact parameter $\beta=r_\mathrm{t}/r_\mathrm{p}$, and collision rates for $10^5$--$10^{10}\;\mathrm{M}_\odot$
    black holes. MS17 refers to MS17.}
    \begin{tabularx}{\textwidth}{lllllllllXXXXXX}
        \hline
        Model & $e$ & $a$ & $i$ & $M$ &  $\Omega$ & $\omega$ & $m_\mathrm{b}$ & $\beta$ & \multicolumn{6}{c}{\hrulefill\  collision fraction for BH ($\mathrm{M}_\odot$)\ \hrulefill} \\
        & & & & & & & & & $10^5$ & $10^6$ & $10^7$ & $10^8$ & $10^9$  & $10^{10}$\\
        & & & & & & & $\mathrm{M}_\odot$ & & $\%$ &$\%$ &$\%$ &$\%$ & $\%$ & $\%$\\ 
        \hline
        1     & $0$      & $1\;\mathrm{au}$      & $0$      & $\mathcal{U}$     & $0$      & $0$ & $1+1$ & $1$    & $6.9$ & $6.9$ & $7.0$ & $6.8$ & $6.8$ & $6.3$  \\
        2     & $\mathcal{G}(0.3, 0.15^2)$     & $1\;\mathrm{au}$      &   0    & $\mathcal{U}$     & $0$      & $0$ & $1+1$ & $1$ & $6.6$ & $6.7$ & $6.8$ & $6.7$ & $6.9$ & $5.4$     \\ 
        3     &  $0$   &  $1\;\mathrm{au}$   & $\cos\left(\mathcal{U}\right)$     & $\mathcal{U}$     & $\mathcal{U}$      & $\mathcal{U}$ & $1+1$ & $1$ & $1.3$ & $1.2$ & $1.2$ & $1.2$ & $1.2$ & $1.2$   \\
        4     &  $\mathcal{G}(0.3, 0.15^2)$     &  $1\;\mathrm{au}$   & $\cos\left(\mathcal{U}\right)$     & $\mathcal{U}$     & $\mathcal{U}$      & $\mathcal{U}$ & $1+1$ & $1$ & $1.2$ & $1.2$ & $1.1$ & $1.1$ & $1.1$ & $1.1$    \\
        5     &  $\mathcal{G}(0.3, 0.15^2)$     & \"Opik's law     &  $\cos\left(\mathcal{U}\right)$     & $\mathcal{U}$     & $\mathcal{U}$      & $\mathcal{U}$ & $1+1$ & $1$ & $14$ & $14$ & $15$ & $15$ & $4.5$ & $1.3$    \\
        6     &  $\mathcal{G}(0.3, 0.15^2)$     & \"Opik's law     &  $\cos\left(\mathcal{U}\right)$     & $\mathcal{U}$     & $\mathcal{U}$      & $\mathcal{U}$ & $10+10$ & $1$   & $14$ & $15$ & $15$ & $15$ & $8.0$ & $2.5$   \\
        7     &  $\mathcal{G}(0.3, 0.15^2)$     & \"Opik's law   &  $\cos\left(\mathcal{U}\right)$     & $\mathcal{U}$     & $\mathcal{U}$      & $\mathcal{U}$ & $10+3$ & $1$ & $14$ & $14$ & $15$ & $16$ & $7.1$ & $2.2$ \\
        8     &  $\mathcal{G}(0.3, 0.15^2)$     & \"Opik's law     &$\cos\left(\mathcal{U}\right)$      & $\mathcal{U}$     & $\mathcal{U}$      & $\mathcal{U}$ & $1+1$ & $\beta^{-2}$  & $6.9$ & $6.9$ & $7.5$ & $6.6$ & $3.7$ & $1.5$   \\
        9     &  $\mathcal{G}(0.3, 0.15^2)$     & \"Opik's law     &$\cos\left(\mathcal{U}\right)$      & $\mathcal{U}$     & $\mathcal{U}$      & $\mathcal{U}$ & KMF & $\beta^{-2}$    & $6.4$ & $6.6$ & $7.1$ & $5.0$ & $2.7$ & $0.84$ \\
        10    & MS17     &  MS17    &   $\cos\left(\mathcal{U}\right)$    &   $\mathcal{U}$   &   $\mathcal{U}$     &  $\mathcal{U}$  & MS17  & $\beta^{-2}$  & $0.27$ & $0.25$ & $0.27$ & $0.26$ & $0.24$ & $0.19$ \\
        \hline
    \end{tabularx}
    \label{tab:all_models}
\end{table*}

\section{Results}
\label{sec:results_6}
We first compare the simulation outcomes across models (Figure~\ref{fig:collisions_cases}). The most common outcome for in-plane, circular-orbit, solar-mass stellar binaries is the Hills mechanism. In Figure~\ref{fig:eject_vel}, we plot the velocities of the escaping star at the point when we stop our simulations, for a $10^6\;\mathrm{M}_\odot$ and a $10^9\;\mathrm{M}_\odot$ black hole, for the scenario where one star is captured. We plot cases 1, 2, 9 and 10 (see Appendix Figure~\ref{fig:data1} for other models).   The velocity distribution implies that these stars would end up as high velocity stars, as expected from the Hills mechanism where one star of the binary is captured by the black hole whereas the other escapes with high velocity \citep{Hills1988}.

To understand the velocity distribution of stars that escape, that is binary escapes undisputed on a hyperbolic orbit, we plot the distribution in Figure~\ref{fig:eject_vel_escape}. The binary escape velocity is slightly lower (factor of $2$) compared to velocity of ejected star.

Changing the eccentricity of the binary does not significantly affect the collision rate, with almost the same number of binaries colliding. As the binary becomes inclined, more models end up captured around the SMBH, with an increase in the number of binaries that escape. Using \"{O}pik's law results in an increase in the number of collisions.
Changing $\beta$ to a distribution $p(\beta) \propto \beta^{-2}$ for a solar-mass binary results in double TDEs for black holes with mass $\leq 10^7\;\mathrm{M}_\odot$, with a small fraction ($< 1\%$) of systems undergoing single stellar TDEs. The collision rate is $\sim 7\%$ across all black hole masses, with the rate decreasing with an increasing mass of the black hole. Generating binary masses from the KMF results in about $6\%$ double TDEs for black holes with mass $\leq 10^6\;\mathrm{M}_\odot$, with about $6\%$ of systems undergoing a TDE of the primary star. Using the MS17 distribution results in collisions for only $\sim 0.2\%$ of systems. Most systems either escape or are captured by the black hole.

Figure~\ref{fig:collision_example} shows two representative simulations of collisions from the MS17 distribution for a $10^9\;\mathrm{M}_\odot$ black hole. The two collision models correspond to binaries with initial semi-major axes of $a = 434\;\mathrm{R}_\odot$ and $a = 21\;\mathrm{R}_\odot$, eccentricities $e = 0.87$ and $e = 0.11$, and impact parameters $\beta = 2.89$ and $\beta = 0.59$, respectively; the stellar masses are $(m_1, m_2) = (0.23, 0.59)\,\mathrm{M}_\odot$ and $(0.63, 0.66)\,\mathrm{M}_\odot$, with differing orbital orientations $(i = 161^\circ, 29^\circ)$, arguments of periapsis $(\omega = 42.08^\circ, 221.37^\circ)$, mean anomalies $(M = 292.36^\circ, 346.71^\circ)$, and longitudes of ascending node $(\Omega = 23.40^\circ, 333.38^\circ)$, illustrating that collisions can arise across a wide range of binary configurations. The von Zeipel-Lidov-Kozai (ZLK) effect occurs in hierarchical three-body systems when the mutual inclination between the inner and outer orbits lies between about $40^\circ$ and $140^\circ$. In this regime, the eccentricity and inclination of the inner orbit undergo coupled oscillations, while the component of its angular momentum along the $z$-axis (defined by the outer orbit) remains conserved \citep{Naoz2016}. Because the systems do not have periods long enough for inner orbits to be affected by the outer orbit, and also because we perform a parabolic orbit, the collision is not a result of ZLK effect.

Figure~\ref{fig:collision_loc} shows the locations of collisions for $10^6\;\mathrm{M}_\odot$ and $10^9\;\mathrm{M}_\odot$ black holes for models 1, 2, 9 and 10 (see Appendix Figure~\ref{fig:data2} for other models). Most collisions occur close to the pericentre. There are gaps in the collision locations for the 1st and 2nd models. These gaps correspond to binary stars narrowly escaping collision and then colliding on the subsequent orbit. Most collisions take place close to the pericentre for all black hole masses; however, for high-mass black holes, collisions can take place before the binary reaches the pericentre. This is due to energy exchange between the inner and outer binary.

To understand at what velocity the two stars collide with each other, we plot the collision velocities in Figure~\ref{fig:collision_vel} for the first four models, where the semi-major axis is $1\;\mathrm{au}$ for a solar-mass stellar binary. For non-inclined models (Models 1 and 2), the collision velocities are tightly clustered around $\sim 617\;\mathrm{km\,s^{-1}}$, which is the escape velocity of a $1+1\,\mathrm{M}_\odot$ binary at contact,
\begin{equation}
    v_{\rm esc} = \sqrt{\frac{2G(m_1+m_2)}{r_1+r_2}} \approx 617\;\mathrm{km\,s^{-1}},
\end{equation}
where $r_1 + r_2 = 2\,\mathrm{R}_\odot$ for a $1\,\mathrm{M}_\odot$ star, using Equation~\ref{eq:stellar_radius}. This implies that most collisions are mergers that is the stars have just enough energy to overcome their mutual gravitational binding energy. For the $10^6\,\mathrm{M}_\odot$ black hole, the distribution is narrow, suggesting that the binary's own internal dynamics dominate the collision velocity. For higher black hole masses, the distribution broadens. This is due to the cross terms in the \textsc{EIH} equation which become more important as the mass of the black hole increases \citep{Sharma2026}. One can see that in Equation~\ref{eq:EIH}, for the pair-wise case, one would ignore the last two terms, which are called the cross-terms, resulting in missing physics.   The gaps in velocity in Model 1 are related to the location of collision, as binary collides on subsequent orbit.   

To test whether collisions prefer head-on or grazing encounters, we next show the impact parameter ($b = |\vec{v}_{12} \times \vec{r}_{12}| / |\vec{v}_{12}|$) in Figure~\ref{fig:collision_b}. For the non-inclined Models 1 and 2, the impact parameter at collision is almost uniformly distributed. For inclined orbits, most collisions occur as grazing encounters, with the collision rate increasing with impact parameter. We also find that most collisions involve prograde binaries, as shown in Figure~\ref{fig:collision_inc}. for models 3, 4, 9 and 10 (Appendix Figure~\ref{fig:data3} shows other models). For binaries on a prograde orbit, the direction of rotation is aligned with the orbital motion of the binary around the black hole. This effect has been noted previously by \citet{Toomre1972} and \citet{Donghia2010}, albeit for the interaction of a galactic disc with another disc modelled as a point mass. The alignment results in more violent encounters due to near-resonance, or the matching of orbital speeds with the peak angular velocity of the companion \citep{Donghia2010}. We find a similar effect here, prograde binaries result in a higher number of collisions. No collisions take place for inclined circular orbits close to retrograde ($\cos i < -0.5$). For the $10^6\;\mathrm{M}_\odot$ black hole, the number of collisions increases as the inclination becomes more aligned with the orbital motion.

To understand the effect of the penetration factor, we plot the $\beta$ values of the models that result in collisions in Figure~\ref{fig:collision_beta}. Most collisions occur at low $\beta$ values, particularly for HMBHs such as the $10^9\;\mathrm{M}_\odot$ black hole. The gravitational force of the black hole on the two stars drives the binary eccentricity to high values during pericentre passage, consistent with the three-body interaction captured by the EIH equations. This effect is more pronounced for tight binaries, which complete more orbits during a single pericentre passage, allowing coherent growth of the binary eccentricity until the stars collide.

\section{Discussion}
\label{sec:discussion_6}

In this work we analysed rates of collisions around different black holes.
The relative velocities at collision are lower than the escape velocity, implying that a significant fraction of these encounters would result in mergers around such black holes. Moreover, collisions can take place at low $\beta \equiv r_\mathrm{t}/r_\mathrm{p}$ values if the binary is tight or eccentric, as gravitational interaction with the black hole drives an increase in eccentricity and eventual collision. Similar to \citet{Mandel2015}, we find that about $6\%$ of binaries would experience single TDEs, with our double TDE fraction about three times lower than their result for Model~9. These numbers differ to due to post-Newtonian effects and because Model~9 uses \"{O}pik's law for the semi-major axis, rather than the distribution derived by \citet{Mandel2015} for semi-major axes that would undergo double TDEs. Our results are consistent with those of \citet{Yu2024}, who studied stellar collisions from binary--SMBH encounters around a $10^6\,\mathrm{M}_\odot$ black hole using the Newtonian code \textsc{Rebound}. They found that in gentle encounters ($\beta \lesssim 1$), the binary survives the pericentre passage but its orbit becomes highly eccentric, leading to a stellar collision with a contact velocity near the escape velocity of the stars, similar to what we find here. 

Our ejection velocities are of the order of magnitude expected from the Hills mechanism,
\begin{equation}
    v_\mathrm{ej} = v_\mathrm{b} \left(\frac{M_\bullet}{m_\mathrm{b}} \right)^{\!\nicefrac{1}{6}}\;,
\end{equation}
where $v_\mathrm{b}$ is the initial orbital velocity of the binary \citep{Hills1988,Bronley2006}. This results in velocities of order $10^{3-4}\;\mathrm{km\,s^{-1}}$, with higher-mass black holes producing higher ejection velocities, as seen in Figure~\ref{fig:eject_vel}. Our results therefore suggest that a population of high velocity stars should exist around HMBHs as well.

\subsection{Rates of collisions}

We calculate the rates of collision by using the empirical relation obtained by \citet{Stone2016} from a sample of $144$ galaxies,
\begin{equation}
    \dot{N} = 2.9\times10^{-5}\;\left(\frac{M_\bullet}{10^8\;\mathrm{M}_\odot} \right)^{B}\left(\frac{a}{\mathrm{R}_\odot} \right)\left(\frac{m_\mathrm{b}}{\mathrm{M}_\odot} \right)^{-1/3} \mathrm{yr^{-1}\;gal^{-1}}\;,
    \label{eq:rate_calc}
\end{equation}
where for a cusp-dominated galaxy such as the Milky Way, $B=-0.223$, and for a core-dominated galaxy, $B=-0.247$. We scaled the \citet{Stone2016} Equation 27 with the semi-major axis and the mass of the binary system, using the tidal radius scaling to obtain Equation~\ref{eq:rate_calc}. The rate of collisions of binary stars can then be estimated as
\begin{equation}
    \dot{N}_\mathrm{coll} = \dot{N} \times f_\mathrm{b} \times f_\mathrm{coll}\;,
\end{equation}
where $f_\mathrm{coll}$ is the fraction of models that undergo collisions in our simulations, each with a total sample of $100{,}000$ models. 

Using Equation~\ref{eq:rate_calc} we determine the collision rate for each setup and then use the median of all data to determine the most likely $\dot{N}$ value for a given distribution. We find collision rates of $\dot{N}_{\rm coll} \sim 10^{-4}\;\mathrm{yr}^{-1}\;\mathrm{gal}^{-1}$ for a $10^6\;\mathrm{M}_\odot$ black hole for our fiducial model, decreasing to $\sim 10^{-5}\;\mathrm{yr}^{-1}\;\mathrm{gal}^{-1}$ for a $10^9\;\mathrm{M}_\odot$ black hole. For the MS17 distribution, the rates are lower by a factor of $10$. This is because the binaries in MS17 are wider compared to the \"Opik's law, resulting in smaller collision fraction.

To estimate the number of observable binary collision events, we multiply our volumetric collision rate by the survey volumes of current optical transient surveys. The volumetric rates are determined in the local universe using the black hole mass function from \citet{Kelly2012}, who derived the mass function of supermassive black holes. We consider number densities in the local universe of $\sim 10^{-2}\,\mathrm{Mpc}^{-3}$ and $\sim 10^{-5}\,\mathrm{Mpc}^{-3}$ for $10^6\;\mathrm{M}_\odot$ and $10^9\;\mathrm{M}_\odot$ black holes, respectively.

For the Zwicky Transient Facility \citep[ZTF;][]{Bellm2019}, which detects transients up to $z \lesssim 0.2$ \citep{Rigault2025}, corresponding to a survey volume of $\sim 3 \times 10^8\;\mathrm{Mpc}^3$, we estimate $\sim 13$ and $4.7$ binary collision events per year for $10^6\;\mathrm{M}_\odot$ black holes for Models~9 and 10. This number decreases with black hole mass, resulting in $\sim 0.5$ and $\sim 0.2$ events per year for black holes of mass $10^8\;\mathrm{M}_\odot$ for Models~9 and 10. In case of $10^9\;\mathrm{M}_\odot$ black holes, for Model~9, and 10, the rates are $\sim0.005$ and $\sim0.004$ events per year. \citet{Wiseman2025} detected $57$ ambiguous nuclear transients (ANTs) using ZTF for SMBH with masses $< 10^8\;\mathrm{M}_\odot$. Our rates from Models 9 and 10 are on lower side.

For the Vera C.\ Rubin Observatory Legacy Survey of Space and Time \citep[LSST;][]{Ivezi2019,Vera2026}, which probes transients out to $z \sim 1$, corresponding to a volume of $\sim 2 \times 10^{11}\,\mathrm{Mpc}^3$, we estimate $\sim 8\mathord,600$ and $3\mathord,000$ events per year for Models~9 and 10 around the $10^6\;\mathrm{M}_\odot$ black hole. For the $10^8\;\mathrm{M}_\odot$ black hole, this would be $\sim330$ and $\sim150$ events per year for Models~9 and 10. For the $10^9\;\mathrm{M}_\odot$ black hole, we predict a detection rate of $3.5$ and $2.8$ events per year for Models~9 and 10. These events would be extremely rare ($10^{-6}$ events per year for LSST) for $10^{10}\;\mathrm{M}_\odot$ black holes, due to the low number densities of such massive black holes in the local universe. 

\subsection{Possible observables}
Based on an extended parameter study, we have classified the outcomes into different categories. Our main focus has been on collisions of binary stars, which might result in observable events and may thus help explain detections of ambiguous nuclear transients \citep{Wiseman2025}. The challenge is that most searches to date look for fast rises in transient light curves. The light curves of TDEs around HMBHs could take years to rise, but still provide a feeding mechanism for the black hole (Sharma et al. in prep.).

 Some binaries undergo the Hills mechanism \citep{Hills1988}, where one star becomes bound to the black hole and the other escapes. The bound stars could provide an explanation for the young stars observed near Sgr~A$^*$ \citep{Ghez1998,Generozov2020,Generozov2021}, implying that even around higher-mass black holes, a population of young stars could exist as a result of the Hills mechanism.

\subsection{Limitations}
Our work has several limitations worth noting. We treat the stars as point particles and therefore cannot model the hydrodynamic outcome of a collision itself --- whether the stars merge, partially disrupt, or eject mass requires smoothed particle hydrodynamics or adaptive grid simulations \citep[e.g.,][]{Yu2025}. We consider only the first pericentre passage of each binary around the black hole. As noted by \citet{Yu2024} and \citet{Antonini2010}, subsequent encounters can further increase the binary eccentricity and hence the collision probability, implying our collision fractions are therefore lower limits. For the rate calculations, we assume a spherical nuclear star cluster (NSC) coexisting with the HMBH and a steady-state loss cone \citep{Stone2016}, neither of which is well established for $M_\bullet \geq 10^8\,\mathrm{M}_\odot$, where relaxation times exceed the Hubble time \citep{Merritt2013}. Our rates should therefore be treated as order-of-magnitude estimates. Finally, we do not model stellar evolution, which may be important for the evolved stellar populations expected near galactic centres.

In this work, we only considered single first-passage orbits around these black holes, but the bound stars could interact with the black hole repeatedly, possibly resulting in partial TDEs or partial quasi-periodic events \citep{Hinkle2024,Nicholl2024}.

For our rates calculations, we assume two-body relaxation is the driving mechanism for the stars getting close to the black hole, and that the galaxy has a spherical geometry with a NSC coexisting with the HMBH. In reality, there is a lack of evidence for the existence of NSCs as the black hole mass increases \citep{Neumayer2020}; hence, the assumption of a spherical distribution of stars around the HMBH may not hold in such environments, as these galaxies have shallower density cusps. \textit{Orbit draining} in non-spherical galaxies could still take place. In axisymmetric potentials, stellar orbits are chaotic outside the sphere of influence. The capture rates of stars differ by a factor of $2$--$3$ between spherical and axisymmetric geometries, but due to relaxation times being closer to the Hubble time for HMBHs, chaotic orbital draining would still fill the loss cone. The flattening of the nucleus would increase the rate of stars entering the loss cone \citep{Merritt2013}. We also assume that about $f_\mathrm{b} \sim 40\%$ of the stars entering the loss cone are binary systems, based on the Galactic Centre population studied by \citet{Gautam2024}. We note that this is a simplification because this fraction only includes young, massive stars, and hence, solar type stars might have a lower fraction, similar to field observations of stellar multiplicity. We also assume full loss cone by using $\beta$ distribution of $\beta^{-2}$, which might not be valid for high mass galaxies which prefer empty loss cone \citep{Stone2016}. 

\section{Conclusion}
\label{sec:conclusion_6}
We performed three-body simulations of binary interactions with black holes of masses ranging from $10^5\;\mathrm{M}_\odot$ to $10^{10}\;\mathrm{M}_\odot$ using the EIH equations. Our findings are as follows:
\begin{enumerate}
    \item Our statistical outcomes are consistent with the results of \citet{Mandel2015} and \citet{Yu2024}. The most common outcome of a binary interacting with the black hole is either escape or capture. The Hills mechanism occurs in all of our models, implying the existence of high velocity stars around HMBHs.

    \item Collisions of binary stars occur predominantly at the pericentre, with typical velocities close to the escape velocity of the binary. For higher-mass black holes, the collision speed extends over a wider range, implying that more binaries undergo mergers.

    \item Both double and single TDEs occur, similar to \citet{Mandel2015}, for our model with the KMF and MS17 binary star distribution. Though MS17 distribution results in order $1\mathord,000$ lower events. 

    \item Most collisions would be detected around lower-mass black holes, with fewer collisions for HMBHs. For LSST, we predict rates of $\sim8\mathord,600$ to $\sim3\mathord,000$ events per year for Models~9 and 10 around $10^6\;\mathrm{M}_\odot$ black holes. We predict $\sim330$ to $\sim150$ collisions per year for Models~9 and 10 around $10^8\;\mathrm{M}_\odot$ black holes. About $\sim3.5$ to $\sim2.8$ events per year might occur for the $10^9\;\mathrm{M}_\odot$ black holes. 
\end{enumerate}

\section*{Acknowledgment}
We thank Ilya Mandel for useful discussions. We also thank Nicholas Stone and Brenna Mockler for helpful comments. This work
was supported by resources awarded under Astronomy Australia Ltd’s ASTAC merit allocation scheme on the OzSTAR and Ngarrgu Tindebeek national
facilities at the Swinburne University of Technology. OzSTAR receives funding from the Australian Government and the Victorian Government. MS thanks Monash University for the Monash International Tuition Scholarship and Monash Graduate Scholarship.  This research was supported by the Australian Research Council (ARC) through Discovery Project DP240103174.  EG acknowledges support from the ARC Discovery Early Career Research Award (DECRA) DE260101802.  

\section*{Data availability}
All data is available at \url{https://dx.doi.org/10.5281/zenodo.22645460}. 
\bibliographystyle{pasa-fixed}
\bibliography{reference}

\appendix
\setcounter{figure}{0}
\renewcommand{\thefigure}
{\thesection.\arabic{figure}}
\setcounter{table}{0}
\renewcommand{\thetable}
{\thesection.\arabic{table}}
\section{Distribution of $\beta$ inside the tidal radius}
\label{app:beta_dis}
To determine the pericentre distribution of stars that enter the tidal radius, we performed a toy model test. We set a $1\;\mathrm{M}_\odot$ star around a $4.3\times10^6\;\mathrm{M}_\odot$ black hole, with an eccentricity of $0.5$. We kick the star on random position on its orbit, with most of the kicks being imparted close to the apocentre, because the star would spend most of its time close to it. We use a kick velocity given by 
\begin{equation}
    \Delta v = \frac{G\; m_*}{b\; v}\;,
\end{equation}
where $m_*$ is assumed to be an external mass of $1\;\mathrm{M_\odot}$, and impact parameter, $b$ is generated from a linear distribution, where minimum and maximum values are $0.5\;\mathrm{au}$ and $6\;\mathrm{au}$, respectively. $v$ is determined by $300\; \mathrm{km/s} \times \sqrt{r_\mathrm{i}/r}$ where $r_\mathrm{i}$ is the starting pericentre of the star of $0.4 \;\mathrm{pc}$. We use $300\;\mathrm{km/s}$ as the dispersion velocity of star at this distance \citep{Pavlik2024}. 
We find that the distribution of stars entering the tidal radius is uniform. Hence, by using change of variable, the probability distribution of $\beta$ would follow
\begin{equation}
    p(\beta) = \beta^{-2}\;.
\end{equation}
This result can be theoretically obtained as well. We use change of variable and definition of $\beta \equiv r_\mathrm{t}/r_\mathrm{p}$, to write 
\begin{align}
    p(\beta) &= p(r_\mathrm{p}) \Bigg|\frac{\mathrm{d}r_\mathrm{p}}{\mathrm{d}\beta}\Bigg|\;\\
    p(\beta) &= p(r_\mathrm{p}) \Bigg|r_\mathrm{r} \frac{\mathrm{d} (1/\beta)}{\mathrm{d}\beta}\Bigg|\;\\
    p(\beta) &\propto \beta^{-2}\;,
\end{align}
where we use $p(r_\mathrm{p}) = \mathrm{constant}$ for a thermal distribution, where using $p(e) = 2 e$, $r_\mathrm{p} = (1-e) a$ and $e \to 1$ as most stars entering the loss-cone are close to the parabolic orbit. We note that the distribution of a binary star entering the tidal radius might not be thermal.

\section{Comparison of EIH with pairwise statistics}
\label{sec:alldata}
\begin{figure}
    \centering
    \includegraphics[width=\columnwidth]{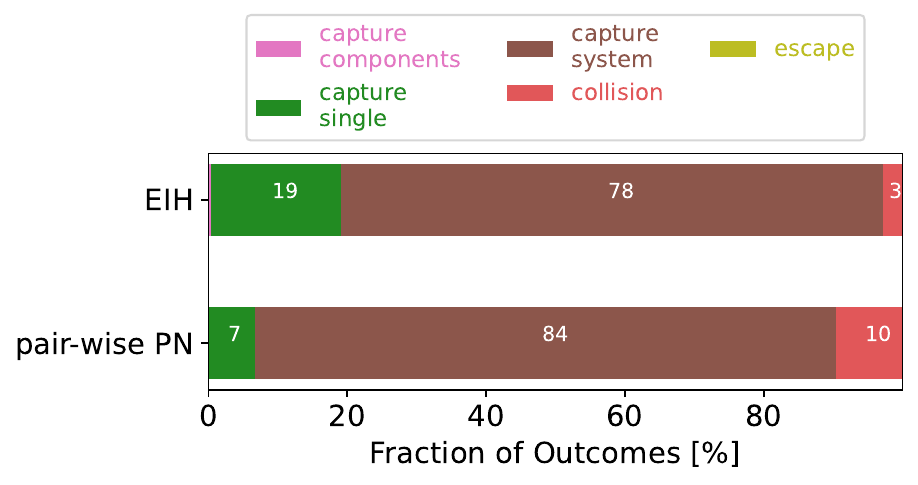}
    \caption{Outcomes for $10^9\;\mathrm{M}_\odot$ black hole, model $9$ for both EIH and pair-wise PN. The latter results in three times more collisions due to the pair-wise method ignoring cross-terms.}
    \label{fig:comparison_eih}
\end{figure}

To understand, how using pair-wise PN affects the statistics compared with EIH, we run $100{,}000$ simulations for Model~9, a $10^9\;\mathrm{M}_\odot$ black hole. Figure~\ref{fig:comparison_eih} shows the rates, showing that pair-wise PN results in more collisions (about three times more) compared with EIH. This is due to the effect caused by missing cross-terms as discussed by \citet{Sharma2026}.

\section{Plots for different distributions}
\begin{figure*}
    \centering
    \includegraphics[width=\textwidth,height=\textheight,keepaspectratio]{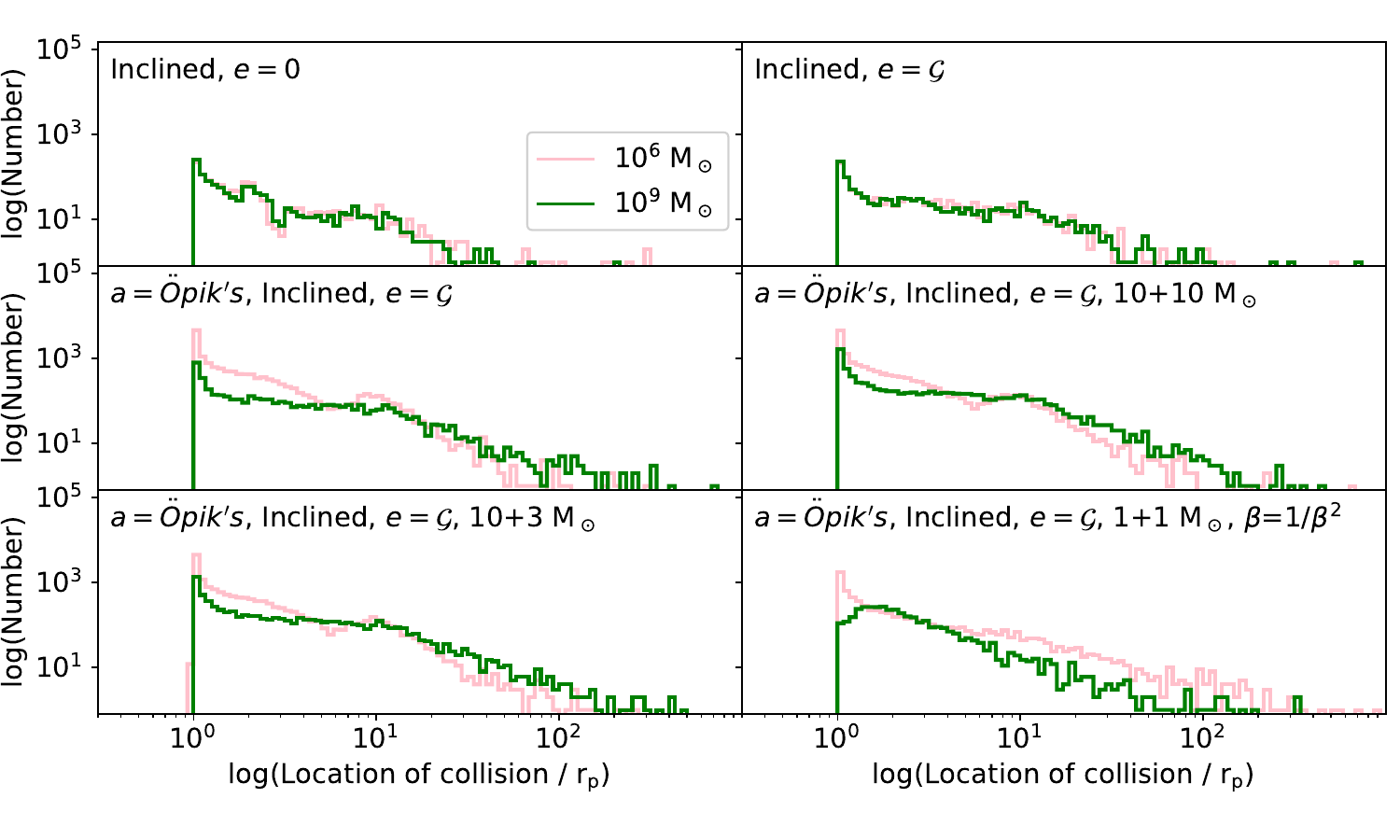}
    \caption{Location of collision for Models 3 to 8.}
    \label{fig:data1}
\end{figure*}

\begin{figure*}
    \centering
    \includegraphics[width=\textwidth,height=\textheight,keepaspectratio]{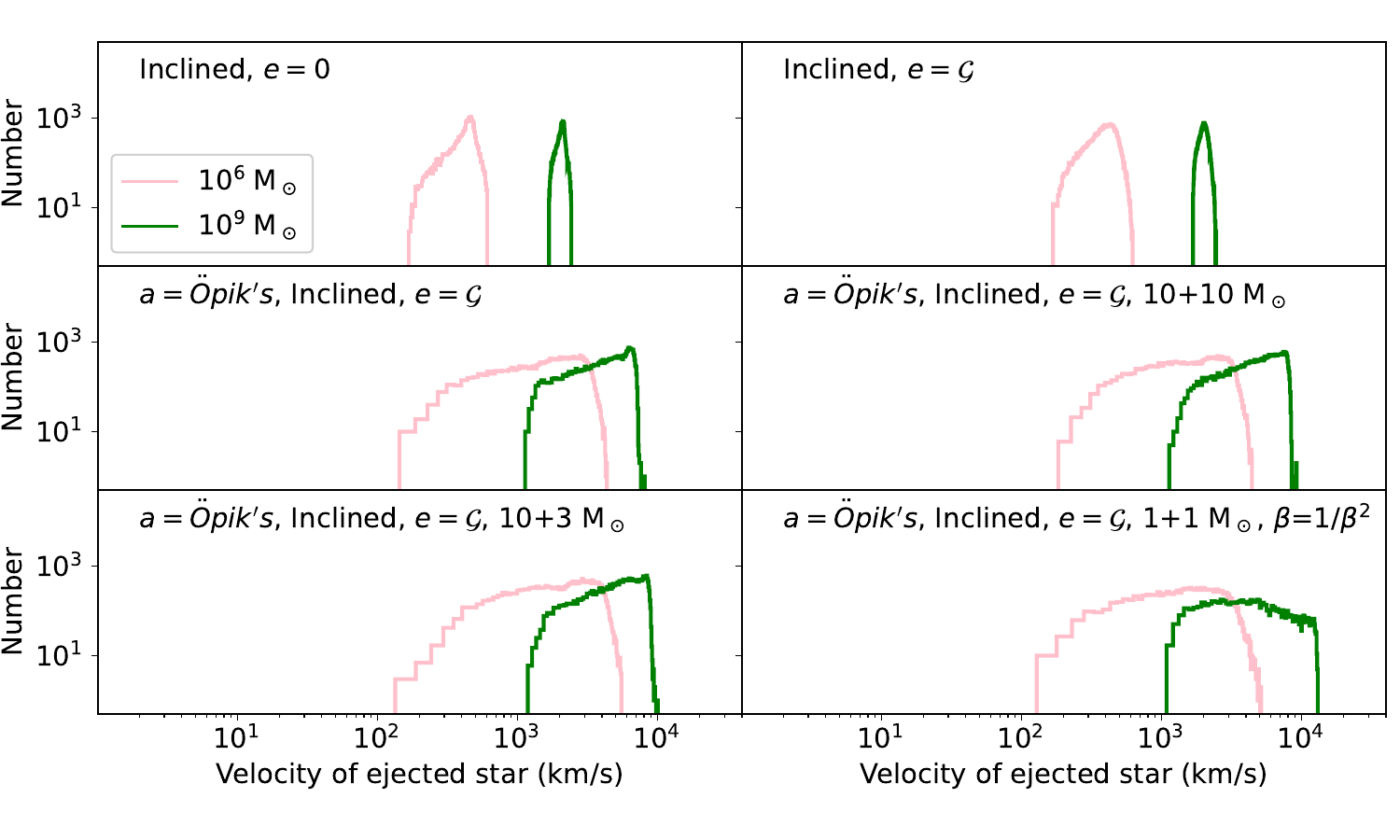}
    \caption{Velocity of ejected star for Models 3 to 8.}
    \label{fig:data2}
\end{figure*}

\begin{figure*}
    \centering
    \includegraphics[width=\textwidth,height=\textheight,keepaspectratio]{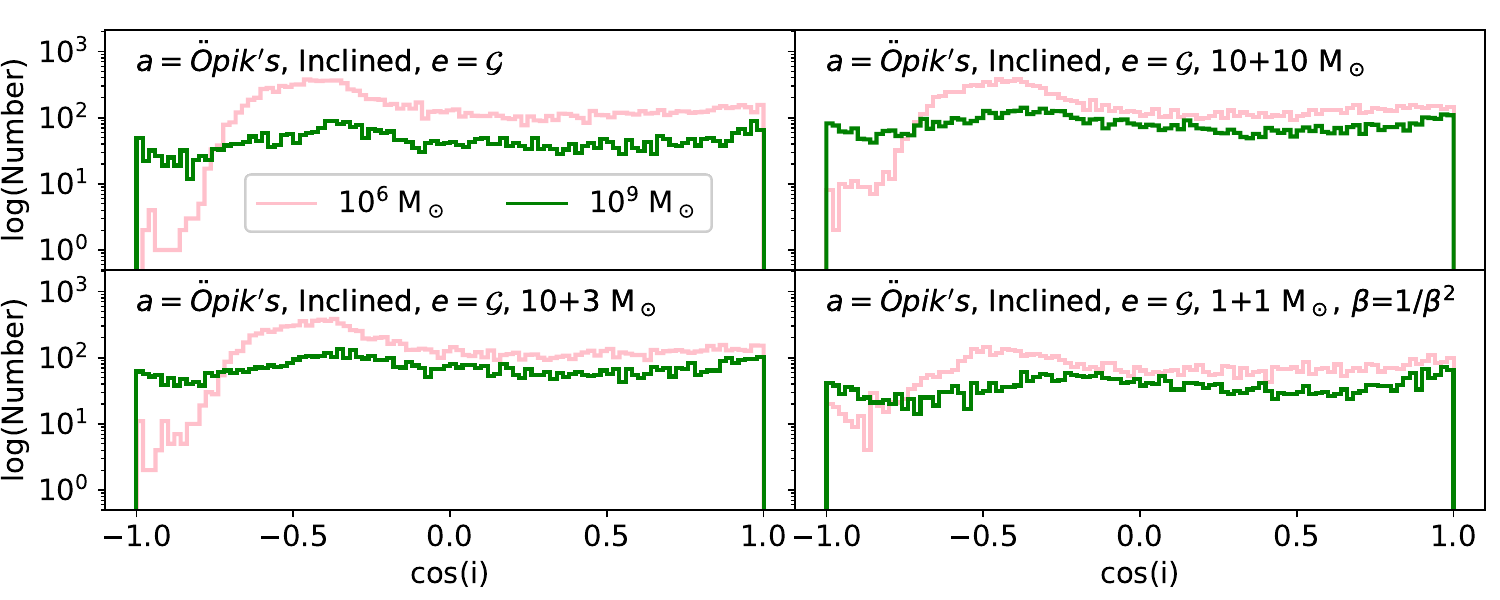}
    \caption{Inclination angle for Models 5 to 8.}
    \label{fig:data3}
\end{figure*}

Figures~\ref{fig:data1} to \ref{fig:data3} show the location of collision for Models 3 to 8, velocity of the ejected star for Models 3 to 8 and inclination angle of the binary for Models 5 to 8, respectively.

\section{Rates of outcomes}

Table~\ref{tab:outcomes} lists the rates of all outcomes across models.

% Requires in preamble: \usepackage{adjustbox}
\begin{table*}
\centering
\caption{Rates of all outcomes for each model for respective black hole mass ($\%$).}
\label{tab:outcomes}
\setlength{\tabcolsep}{1.5pt}
\footnotesize
\begin{adjustbox}{max width=\textwidth, max totalheight=0.9\textheight}
\begin{tabular}{l|cccccc|cccccc|cccccc}
\hline
% ================= Block 1: Models 1-3 =================
\multicolumn{1}{c|}{} & \multicolumn{6}{c|}{Model 1} & \multicolumn{6}{c|}{Model 2} & \multicolumn{6}{c}{Model 3} \\
\hline
\multicolumn{1}{c|}{} & \multicolumn{6}{c|}{Black hole mass ($\mathrm{M}_\odot$)} & \multicolumn{6}{c|}{Black hole mass ($\mathrm{M}_\odot$)} & \multicolumn{6}{c}{Black hole mass ($\mathrm{M}_\odot$)} \\
Outcomes & $10^5$ & $10^6$ & $10^7$ & $10^8$ & $10^9$ & $10^{10}$ & $10^5$ & $10^6$ & $10^7$ & $10^8$ & $10^9$ & $10^{10}$ & $10^5$ & $10^6$ & $10^7$ & $10^8$ & $10^9$ & $10^{10}$ \\
\hline
TDE both           & 0 & 0 & 0 & 0 & 0 & 0 & 0 & 0 & 0 & 0 & 0 & 0 & 0 & 0 & 0 & 0 & 0 & 0 \\
TDE single         & 0 & 0 & 0 & 0 & 0 & 0 & 0 & 0 & 0 & 0 & 0 & 0 & 0 & 0 & 0 & 0 & 0 & 0 \\
Capture components & 0 & 0 & 0 & 0 & 0 & 0 & 0 & 0 & 0 & 0 & 0 & 0 & $2\times10^{-3}$ & 0 & 0 & 0 & 0 & 0 \\
Capture single     & 89.97 & 90.16 & 90.02 & 90.32 & 90.34 & 91.24 & 89.85 & 89.67 & 89.52 & 89.83 & 89.71 & 91.89 & 55.26 & 55.16 & 55.87 & 55.70 & 55.76 & 56.11 \\
Capture system     & 0.54 & 0.49 & 0.49 & 0.54 & 1.29 & 2.48 & 0.72 & 0.77 & 0.71 & 0.73 & 1.62 & 2.68 & 26.21 & 26.08 & 25.77 & 27.71 & 40.03 & 42.75 \\
Escape             & 2.60 & 2.44 & 2.54 & 2.34 & 1.57 & 0 & 2.81 & 2.87 & 2.97 & 2.77 & 1.79 & 0 & 17.25 & 17.05 & 17.17 & 15.39 & 3.02 & 0 \\
Collision          & 6.87 & 6.90 & 6.95 & 6.79 & 6.79 & 6.28 & 6.61 & 6.69 & 6.79 & 6.67 & 6.87 & 5.42 & 1.27 & 1.23 & 1.19 & 1.19 & 1.18 & 1.15 \\
\hline
% ================= Block 2: Models 4-6 =================
\multicolumn{1}{c|}{} & \multicolumn{6}{c|}{Model 4} & \multicolumn{6}{c|}{Model 5} & \multicolumn{6}{c}{Model 6} \\
\hline
\multicolumn{1}{c|}{} & \multicolumn{6}{c|}{Black hole mass ($\mathrm{M}_\odot$)} & \multicolumn{6}{c|}{Black hole mass ($\mathrm{M}_\odot$)} & \multicolumn{6}{c}{Black hole mass ($\mathrm{M}_\odot$)} \\
Outcomes & $10^5$ & $10^6$ & $10^7$ & $10^8$ & $10^9$ & $10^{10}$ & $10^5$ & $10^6$ & $10^7$ & $10^8$ & $10^9$ & $10^{10}$ & $10^5$ & $10^6$ & $10^7$ & $10^8$ & $10^9$ & $10^{10}$ \\
\hline
TDE both           & 0 & 0 & 0 & 0 & 0 & 0 & 0 & 0 & 0 & 0 & 0 & 0 & 0 & 0 & 0 & 0 & 0 & 0 \\
TDE single         & 0 & 0 & 0 & 0 & 0 & 0 & 0 & 0 & 0 & 0 & 0 & 0 & 0 & 0 & 0 & 0 & 0 & 0 \\
Capture components & $10^{-3}$ & 0 & 0 & 0 & 0 & 0 & $10^{-3}$ & 0 & 0 & 0 & 0 & 0 & $6\times10^{-3}$ & $2\times10^{-3}$ & 0 & 0 & 0 & 0 \\
Capture single     & 56.69 & 57.34 & 57.39 & 57.55 & 58.84 & 61.17 & 52.54 & 52.47 & 52.25 & 55.06 & 61.08 & 62.68 & 52.63 & 52.66 & 52.77 & 53.54 & 60.53 & 62.46 \\
Capture system     & 25.73 & 25.08 & 25.02 & 26.16 & 36.98 & 37.72 & 21.99 & 22.33 & 23.85 & 28.13 & 34.06 & 36.03 & 21.71 & 21.62 & 21.72 & 24.40 & 29.33 & 35.02 \\
Escape             & 16.37 & 16.43 & 16.45 & 15.15 & 3.04 & 0 & 11.64 & 11.11 & 9.04 & 1.84 & 0.40 & 0 & 11.53 & 11.15 & 10.69 & 6.74 & 2.15 & 0.06 \\
Collision          & 1.21 & 1.15 & 1.14 & 1.14 & 1.14 & 1.09 & 13.82 & 14.09 & 14.86 & 14.97 & 4.46 & 1.28 & 14.49 & 14.56 & 14.81 & 15.32 & 7.99 & 2.45 \\
\hline
% ================= Block 3: Models 7-9 =================
\multicolumn{1}{c|}{} & \multicolumn{6}{c|}{Model 7} & \multicolumn{6}{c|}{Model 8} & \multicolumn{6}{c}{Model 9} \\
\hline
\multicolumn{1}{c|}{} & \multicolumn{6}{c|}{Black hole mass ($\mathrm{M}_\odot$)} & \multicolumn{6}{c|}{Black hole mass ($\mathrm{M}_\odot$)} & \multicolumn{6}{c}{Black hole mass ($\mathrm{M}_\odot$)} \\
Outcomes & $10^5$ & $10^6$ & $10^7$ & $10^8$ & $10^9$ & $10^{10}$ & $10^5$ & $10^6$ & $10^7$ & $10^8$ & $10^9$ & $10^{10}$ & $10^5$ & $10^6$ & $10^7$ & $10^8$ & $10^9$ & $10^{10}$ \\
\hline
TDE both           & 0 & 0 & 0 & 0 & 0 & 0 & 9.05 & 9.32 & 6.51 & 0 & 0 & 0 & 6.04 & 6.11 & 0.56 & $10^{-3}$ & 0 & 0 \\
TDE single         & 0 & 0 & 0 & 0 & 0 & 0 & 0.86 & 0.43 & 0.10 & 0 & 0 & 0 & 6.22 & 6.09 & 0.77 & $10^{-3}$ & 0 & 0 \\
Capture components & $4\times10^{-3}$ & 0 & 0 & 0 & 0 & $3\times10^{-3}$ & 0.58 & 0.51 & 0.24 & 0 & 0 & 0.39 & 0.49 & 0.42 & 0.04 & $2\times10^{-3}$ & 0.38 & 1.38 \\
Capture single     & 52.70 & 52.58 & 52.70 & 53.44 & 60.64 & 62.39 & 32.70 & 32.86 & 34.95 & 29.80 & 19.57 & 20.03 & 30.75 & 31.01 & 34.13 & 22.95 & 18.57 & 18.85 \\
Capture system     & 21.59 & 22.13 & 22.05 & 26.35 & 31.19 & 35.46 & 33.77 & 37.83 & 43.66 & 61.31 & 76.43 & 78.02 & 36.54 & 41.45 & 54.04 & 71.37 & 78.28 & 78.91 \\
Escape             & 11.34 & 11.06 & 10.64 & 4.41 & 1.05 & 0 & 16.13 & 12.17 & 7.05 & 2.27 & 0.26 & 0.02 & 13.56 & 8.37 & 3.38 & 0.65 & 0.07 & $9\times10^{-3}$ \\
Collision          & 14.36 & 14.23 & 14.61 & 15.78 & 7.11 & 2.15 & 6.91 & 6.88 & 7.48 & 6.62 & 3.73 & 1.53 & 6.39 & 6.58 & 7.12 & 5.02 & 2.65 & 0.84 \\
\hline
% ================= Block 4: Model 10 =================
% Lines in this block use \cline{1-7} so they stop at the Model 10 columns.
\multicolumn{1}{c|}{} & \multicolumn{6}{c|}{Model 10} & \multicolumn{12}{c}{} \\
\cline{1-7}
\multicolumn{1}{c|}{} & \multicolumn{6}{c|}{Black hole mass ($\mathrm{M}_\odot$)} & \multicolumn{12}{c}{} \\
Outcomes & $10^5$ & $10^6$ & $10^7$ & $10^8$ & $10^9$ & $10^{10}$ & \multicolumn{12}{c}{} \\
\cline{1-7}
TDE both           & $8\times10^{-3}$ & $5\times10^{-3}$ & $10^{-2}$ & 0 & 0 & 0 & \multicolumn{12}{c}{} \\
TDE single         & $8\times10^{-3}$ & $9\times10^{-3}$ & $9\times10^{-3}$ & 0 & 0 & 0 & \multicolumn{12}{c}{} \\
Capture components & $10^{-3}$ & $3\times10^{-3}$ & $2\times10^{-3}$ & 0 & 0 & 0.04 & \multicolumn{12}{c}{} \\
Capture single     & 13.02 & 12.74 & 12.96 & 12.91 & 13.06 & 13.71 & \multicolumn{12}{c}{} \\
Capture system     & 26.87 & 29.59 & 35.71 & 45.19 & 58.18 & 69.78 & \multicolumn{12}{c}{} \\
Escape             & 59.82 & 57.40 & 51.03 & 41.64 & 298.51 & 16.28 & \multicolumn{12}{c}{} \\
Collision          & 0.27 & 0.25 & 0.27 & 0.26 & 0.24 & 0.19 & \multicolumn{12}{c}{} \\
\cline{1-7}
\end{tabular}
\end{adjustbox}
\end{table*}

\end{document}